\documentclass[twocolumn]{aastex631}

\usepackage{xspace}

\newcommand\hi{\mbox{\sc Hi}}

\newcommand\kms{km s$^{-1}$}
\newcommand\htwo{H$_2$\xspace}

\received{June 12, 2024}
\revised{July 17, 2024}
\accepted{July 18, 2024}

\shorttitle{LGLBS: First Localized CNM Properties in NGC 6822}
\shortauthors{Pingel et al.}
\graphicspath{{./}{figures/}}

\begin{document}

\title{The Local Group L-Band Survey: The First Measurements of Localized Cold Neutral Medium Properties in the Low-Metallicity Dwarf Galaxy NGC 6822}

\correspondingauthor{Nickolas M. Pingel}
\email{nmpingel@wisc.edu}

\author[0000-0001-9504-7386]{Nickolas~M.~Pingel}
\affiliation{University of Wisconsin–Madison, Department of Astronomy, 475 N Charter St, Madison, WI 53703, USA}

\author[0009-0005-1781-5665]{Hongxing Chen}
\affiliation{University of Wisconsin–Madison, Department of Astronomy, 475 N Charter St, Madison, WI 53703, USA}

\author[0000-0002-3418-7817]{Sne{\v z}ana Stanimirovi{\'c}}
\affiliation{University of Wisconsin–Madison, Department of Astronomy, 475 N Charter St, Madison, WI 53703, USA}

\author[0000-0001-9605-780X]{Eric~W.~Koch}
\affiliation{Center for Astrophysics $\mid$ Harvard \& Smithsonian, 60 Garden St., 02138 Cambridge, MA, USA}
\author[0000-0002-2545-1700]{Adam~K.~Leroy}
\affiliation{Department of Physics, The Ohio State University, Columbus, Ohio 43210, USA}
\affiliation{Center for Cosmology \& Astro-Particle Physics, The Ohio State University, Columbus, Ohio 43210, USA}

\author[0000-0002-5204-2259]{Erik Rosolowsky}
\affiliation{Dept. of Physics, University of Alberta, 4-183 CCIS, Edmonton, Alberta, T6G 2E1, Canada}

\author[0000-0003-2896-3725]{Chang-Goo Kim}
\affiliation{Department of Astrophysical Sciences, Princeton University, 4 Ivy Lane, Princeton, NJ 0844, USA}

\author[0000-0002-1264-2006]{Julianne J.~Dalcanton}
\affiliation{Center for Computational Astrophysics, Flatiron Institute, 162 Fifth Avenue, New York, NY 10010, USA}
\affiliation{Department of Astronomy, Box 351580, University of Washington, Seattle, WA 98195, USA}

\author[0000-0003-4793-7880]{Fabian Walter}
\affiliation{Max-Planck-Institut f\"ur Astronomie, K\"onigstuhl 17, 69117 Heidelberg, Germany
}

\author[0000-0003-4961-6511]{Michael~P.~Busch}
\affiliation{Department of Astronomy \& Astrophysics, University of California, San Diego, 9500 Gilman Drive, La Jolla, CA 92093, USA}

\author[0000-0001-8241-7704]{Ryan Chown}
\affiliation{Department of Physics, The Ohio State University, Columbus, Ohio 43210, USA}
\affiliation{Center for Cosmology \& Astro-Particle Physics, The Ohio State University, Columbus, Ohio 43210, USA}

\author[0000-0002-3106-7676]{Jennifer Donovan Meyer}
\affiliation{National Radio Astronomy Observatory, 520 Edgemont Rd., Charlottesville, VA 22968, USA}

\author[0000-0002-1185-2810]{Cosima Eibensteiner}
\altaffiliation{Jansky Fellow of the National Radio Astronomy Observatory}
\affiliation{National Radio Astronomy Observatory, 520 Edgemont Road, Charlottesville, VA 22903, USA}

\author[0000-0002-3322-9798]{Deidre~A.~Hunter}
\affiliation{Lowell Observatory, 1400 W Mars Hill Road, Flagstaff, AZ 86001, USA}

\author[0000-0002-4781-7291]{Sumit K. Sarbadhicary}
\affiliation{Department of Physics, The Ohio State University, Columbus, Ohio 43210, USA}
\affiliation{Center for Cosmology \& Astro-Particle Physics, The Ohio State University, Columbus, Ohio 43210, USA}

\author[0000-0003-1356-1096]{Elizabeth Tarantino}
\affiliation{Space Telescope Science Institute, 3700 San Martin Drive, Baltimore, MD 21218, USA}

\author[0000-0002-5877-379X]{Vicente Villanueva}
\affiliation{Departamento de Astronom{\'i}a, Universidad de Concepci{\'o}n, Barrio Universitario, Concepci{\'o}n, Chile}

\author[0000-0002-0012-2142]{Thomas~G.~Williams}
\affiliation{Sub-department of Astrophysics, Department of Physics, University of Oxford, Keble Road, Oxford OX1 3RH, UK}



\begin{abstract}
Measuring the properties of the cold neutral medium (CNM) in low-metallicity galaxies provides insight into heating and cooling mechanisms in early Universe-like environments. We report detections of two localized atomic neutral hydrogen (\hi) absorption features in NGC 6822, a low-metallicity (0.2 Z$_{\odot}$) dwarf galaxy in the Local Group. These are the first unambiguous CNM detections in a low-metallicity dwarf galaxy outside the Magellanic Clouds. The Local Group L-Band Survey (LGLBS) enabled these detections due to its high spatial (15 pc for \hi\ emission) and spectral (0.4 \kms) resolution. We introduce LGLBS and describe a custom pipeline to search for \hi\ absorption at high angular resolution and extract associated \hi\ emission. A detailed Gaussian decomposition and radiative transfer analysis of the NGC 6822 detections reveals five CNM components, with key properties: a mean spin temperature of 32$\pm$6 K, a mean CNM column density of 3.1$\times$10$^{20}$ cm$^{-2}$, and CNM mass fractions of 0.33 and 0.12 for the two sightlines. Stacking non-detections does not reveal low-level signals below our median optical depth sensitivity of 0.05. One detection intercepts a star-forming region, with the \hi\ absorption profile encompassing the CO (2$-$1) emission, indicating coincident molecular gas and a depression in high-resolution \hi\ emission. We also analyze a nearby sightline with deep, narrow \hi\ self-absorption dips, where the background warm neutral medium is attenuated by intervening CNM. The association of CNM, CO, and H$\alpha$ emissions suggests a close link between the colder, denser \hi\ phase and star formation in NGC 6822. 
\end{abstract}

\vspace{-19mm}

\section{Introduction} \label{sec:intro}
Atomic hydrogen (\hi) gas is the primary component of the interstellar medium (ISM) by mass in most galaxies and makes up the majority of the baryonic mass in many low-mass galaxies. It is used to form molecular hydrogen (H$_2$), and has a crucial role in the shielding of dense molecular gas, where stars form, from photodissociation processes \citep{mcclure-griffiths2023}. How giant molecular clouds (GMCs) form out of the diffuse \hi\ is still not well understood, the \hi\ gas is considered as the main formation reservoir of GMCs \citep{shu1973, blitz2007, kim2006, audit2005, heitsch2005, clark2012, dobbs2014}. 

Theoretical models of steady-state ISM heating and cooling predict the nature of the \hi\ thermal structure. In the Solar neighborhood, two thermally stable phases --- the cold neutral
medium (CNM) and warm neutral medium (WNM) --- are expected, with the kinetic temperature ($T_{\rm k})$ being mainly in the range 25 K to 250 K and 5000 K to 8000 K, respectively. These steady-state models demonstrate that \hi\ in the range of about 250 K$\leq T_{\rm k}\leq 5000$ K is thermally unstable and short-lived, although this depends on the details of the local heating and cooling processes \citep{field1969,wolfire1995, wolfire2003, bialy2019}. The local conditions of the ISM influence the properties of distinct \hi\ phases. For example, in metal-poor gas with low dust content, the photoelectric heating from the surface of dust grains and polycyclic aromatic hydrocarbons (PAHs) is less effective \citep{bialy2019}. As a result, at low metallicity one might expect that the CNM should be colder. Additionally, when the UV radiation field increases by a factor of 10, the thermal equilibrium curve, which markes where thermally-stable CNM and WNM can co-exist, shifts to higher pressures \citep{mcclure-griffiths2023}. 

The CNM has low spin temperatures and high optical depths, which influence the emission at small scales ($\lesssim$ 100 pc) at least within the Milky Way. For example, \citet{Heiles2003} observed examples of narrow \hi\ emission features which are also seen in \hi\ absorption at high Galactic latitudes. Similarly, \hi\ self-absorption in the Milky Way produces characteristic line profiles that can be used to extract information about the properties of the CNM (e.g., \citealt{gibson2005, soler2020, wang2020}). However, in other galaxies, the interpretation of \hi\ emission profiles is usually complicated by limitations in spatial/spectral resolution of the radio telescope and overall degeneracy in the line broadening caused by turbulence and the blending of emission components arising from velocity crowding and a mix of thermal phases \citep{koch2021}. 
Given this, 
the most direct way of detecting the CNM
is through \hi\ 21 cm absorption against strong radio background sources (e.g., \citealt{murray2018, jameson2019, dempsey2022}). Such measurements enable constraints of the spin temperature and optical depth when the associated \hi\ emission spectrum is also considered. And because collisions are sufficient to populate the energy levels at the volume densities of the CNM (10 cm$^{-3}-$100 cm$^{-3}$), a measurement of the spin temperature becomes a useful proxy for $T_{k}$ \citep{field1958, Liszt2001}.

Numerous studies have focused on sightlines through the local Milky Way disk to study the \hi\ emission and absorption in conjunction with one another at high enough spectral and angular resolution to provide constraints on CNM properties \citep{Heiles2003, dickey2009, roy2013, murray2018}. Recent sensitive wide-field observations enabled by the Australian Telescope Compact Array (ATCA) and the Australian Square Kilometre Array Pathfinder (ASKAP) have provided the first statistical samples of the \hi\ thermal structure at low-metallicity by detecting \hi\ absorption in the direction of background radio continuum sources behind the Magellanic Clouds \citep{dickey2000, marx-zimmer2000, jameson2019, dempsey2022}. But due to limitations in sensitivity, there have only been a handful of \hi\ absorption detections with associated \hi\ emission in external galaxies outside the Magellanic System, specifically M31 and M33 \citep{dickey1988, braun1992, dickey1993}. A main goal of extragalactic \hi\ surveys is to probe the properties and abundance of the CNM in galaxies with properties distinct from the Milky Way, but so far there have been no detections of CNM absorption in any low metallicity galaxy beyond the LMC and SMC.

A new major survey called the Local Group L-Band Survey (LGLBS), using $\sim$1700~hours of observing time with the post-upgrade Karl G.\ Jansky Expanded Very Large Array (VLA), combines high spatial and spectral resolution line and continuum observations of the six star-forming Local Group galaxies visible to the VLA to probe a wide parameter space in terms of astrophysical environments (E.~Koch et al., in preparation; N.~Pingel et al, in preparation). With the advent of the LGLBS, we now have the spectral resolution, angular resolution, and sensitivity to detect 21-cm absorption against bright background or in-galaxy continuum sources outside the Magellanic System for the first time. 

This paper presents the first results of this effort and is organized as follows: Section~\ref{sec:previous} provides a brief overview of the findings from previous studies of NGC 6822, Section~\ref{sec:survey_obs} a summary of LGLBS, and outlines our methods for the calibration, and extraction of the absorption and emission spectra; Section~\ref{sec:results} discusses the characteristics of our detections; Section~\ref{sec:indv_CNM} summarizes the joint decomposition analysis of the emission and absorption spectra to measure physical properties; Section~\ref{sec:multiwavelength_comparison} explores the spatial and spectral association of the \hi\ absorptoin with other tracers; Section~\ref{sec:discussion} places the CNM properties in NGC 6822 in context with previous studies of the high-latitude Milky Way and Magellanic System and discusses the association with nearby molecular gas; finally, we discuss future work and summarize in Section~\ref{sec:conculsions}.

\section{Previous Studies of Cold Gas in NGC 6822}\label{sec:previous}

At a distance of $480$~kpc, NGC 6822 is the nearest actively star forming low-metallicity galaxy beyond the SMC. In many ways, NGC 6822 resembles the SMC. For example, it has metallicity $12+\log_{10} {\rm (O/H)} = 8.02\pm0.05$ \citep{garcia2016}, similar to that of the SMC (8.01$\pm$0.02). It is also similarly gas-rich with an atomic mass of $1.3\times10^8$ M$_{\odot}$ \citep{deBlok2006}. 
However, NGC 6822 has a star formation rate (SFR) of 0.015 M$_{\odot}$ yr$^{-1}$ \citep{efremova2011}, which is roughly two times lower than that of the SMC \citep{bolatto2011}, and a global gas-to-dust ratio of $\sim480$ \citep{schruba2017}.

 Most of the star formation in NGC 6822 is occurring in four massive cloud complexes \citep{deBlok2006b}. These together host about $2/3$ of the total star formation activity in the galaxy, have each been forming stars for at least $\sim 10$~Myr and host $> 100$ OB stars \citep{schruba2017}. Because they represent key targets, these regions already have high resolution, high sensitivity multi-wavelength data sets \citep{cormier2014, lenkic2024, nally2024}. More recently, the CO mapping from ALMA presented in \cite{schruba2017} reveals the presence of $2-3$~pc sized CO cores scattered across each region. Determining the density and kinematic properties of the CNM that is associated with these dense, star-forming CO structures will provide insights into the physics of star forming regions in low-metallicity environments, as well as studying the association with young stellar objects from \textit{Spitzer} \citep{hirschauer2020} and the JWST \citep{lenkic2024}.

\citet{deBlok2000}, \citet{park2022}, and \citet{namumba2017} all observed and analyzed \hi\ emission from 6822. These studies considered the existence of narrow line \hi\ components in the emission profiles through Gaussian decomposition techniques and the implications for gas stability but so far no \hi\ has been detected in absorption. Thus, there has been no unambiguous measurement of the abundance and properties of the cold atomic gas in this galaxy.

\section{Observations and Data Processing}\label{sec:survey_obs}

\subsection{The Local Group L-band Survey (LGLBS)}
\label{subsec:survey_overview}
LGLBS is an extra-large project that observed L-band line and continuum emission from the Local Group spirals M31 and M33 and the dwarf galaxies IC~10, IC~1613, NGC~6822, and WLM. The observations combine all available VLA configurations using a 1:1:1:0.5 ratio of integration time between A, B, C, and D configurations. The full survey is described in E. Koch et al. (in preparation) and the 21-cm imaging in N. Pingel et al. (in preparation). The spectral setup covers the 21-cm line at high velocity resolution, the four 18 cm OH lines, and the polarized 1{-}2~GHz radio continuum. 

For NGC~6822, we covered the 21-cm line with an 8 MHz ($1700$~km~s$^{-1}$) spectral window with 1.95~kHz resolution ($0.4$~km~s$^{-1}$) and dual polarization. The bandwidth of 1700 km~s$^{-1}$ allows for effective continuum subtraction, a large search space for faint, extended \hi, and fully captures the Milky Way \hi\ foreground. The $0.4$~km~s$^{-1}$ velocity resolution ensures that we can resolve the line profile down to a $\sim 100$~K line (line width $\sim 0.8$~km~s$^{-1}$). We covered the \hi\ disk of the galaxy using three fields, which were observed using integration times of 11~hr per field in each of the A, B, and C configurations, and 5.5~hr per field in the D configuration. All VLA 21-cm \hi\ observations obtained previously with the WIDAR correlator are also incorporated here (project IDs: 13A-213, 14B-212, and 20A-346).

\subsection{Calibration and Processing}
\label{subsec:calibration}

The full pipeline processing of LGLBS is described in E. Koch. et al. (in preparation) and the 21-cm imaging in N. Pingel et al. (in preparation). Briefly, we calibrate the data using the standard VLA pipeline (version 6.4.1) with additional modifications appropriate for spectral line observations. Each observation is manually quality assured using products created with our custom software \textsc{QAPlotter\footnote{\url{https://github.com/LocalGroup-VLALegacy/QAPlotter}}}, which produces interactive summary plots of the visibilities and coarse-resolution dirty science images of each spectral line. We apply the manual flags generated by this process and then iteratively re-run the pipeline until the calibrated visibilities are sufficient science quality.

The \hi~emission data are imaged using a custom pipeline described in detail by N. Pingel et al. (in preparation). This pipeline is designed to be run on the resources available through the Center for High Throughput Computing (CHTC) at the University of Wisconsin$-$Madison. To take full advantage of the $\sim$20,000 available computing cores through the CHTC, our imaging pipeline takes a directed acyclic graph (DAG) approach, where relationships and dependencies are established between subsequent steps in the processing. The DAG approach is useful in workflows where jobs must be run in a particular order without the need to cycle back to previous processing steps. In other words, the steps in a workflow are distinct, depend on one another, and will never form a closed loop. DAGMan, the builtin workflow manager within the HTConder scheduling software on the CHTC, monitors the status of jobs associated with a specific step in the workflow and will automatically submit jobs linked to subsequent step in the work flow to ensure the workflow is executed in the correct order. For example, an early step in the pipeline submits simultaneous jobs to run the task \textsc{mstransform} on the measurement sets from each VLA track to ensure each possess a uniform spectral axis. The DAG approach verifies each of these jobs has successfully completed before moving on to subsequent processing steps, limiting the need for human interaction during the imaging stage. This pipeline utilizes several tasks available through the Common Astronomy Software Application (CASA; \citealt{bean2022_casa}) package (version 6.5.0).

As part of the \hi\ emission processing, we perform a visibility-based continuum subtraction. For each measurement set, a first-order polynomial is fit across frequency, excluding channels corresponding to LSR velocities $-150$\kms~to 200 \kms\ to select emission-free channels around NGC 6822 and the Milky Way foreground. The deconvolution step is efficiently distributed across the cluster by splitting out each individual 0.4~km s$^{-1}$ spectral channels to be imaged in individual jobs that run in parallel before recombining into the final cube. During deconvolution, we use the briggs weighting scheme \citep{briggs1995} with the robustness set to 1.0 to get the best compromise between the angular resolution, noise, and sidelobe levels, and a spatial pixel size of 0.75$''\times$0.75$''$. 

The final beam size of the \hi\ emission data cube is 7.0$''\times$5.2$''$, or $\approx 12 \times 17$~pc at the 480~kpc distance to NGC 6822, with a position angle of the major axis from the North direction of $-$3.0 deg. The rms noise is 7 K per 0.4 km $^{-1}$ channel, providing a 5$\sigma$ detection limit of 1.8$\times$10$^{20}$ cm$^{-2}$ over a 20 km s$^{-1}$ line for the $\hi$ column density, under the assumption that the emission is optically thin. This is similar to the ATCA observations of \cite{deBlok2000} while also resolving physical scales 10 times smaller. The spectral axis of the final cube is converted to the LSR reference frame. 

\begin{figure*}
\centering
\includegraphics{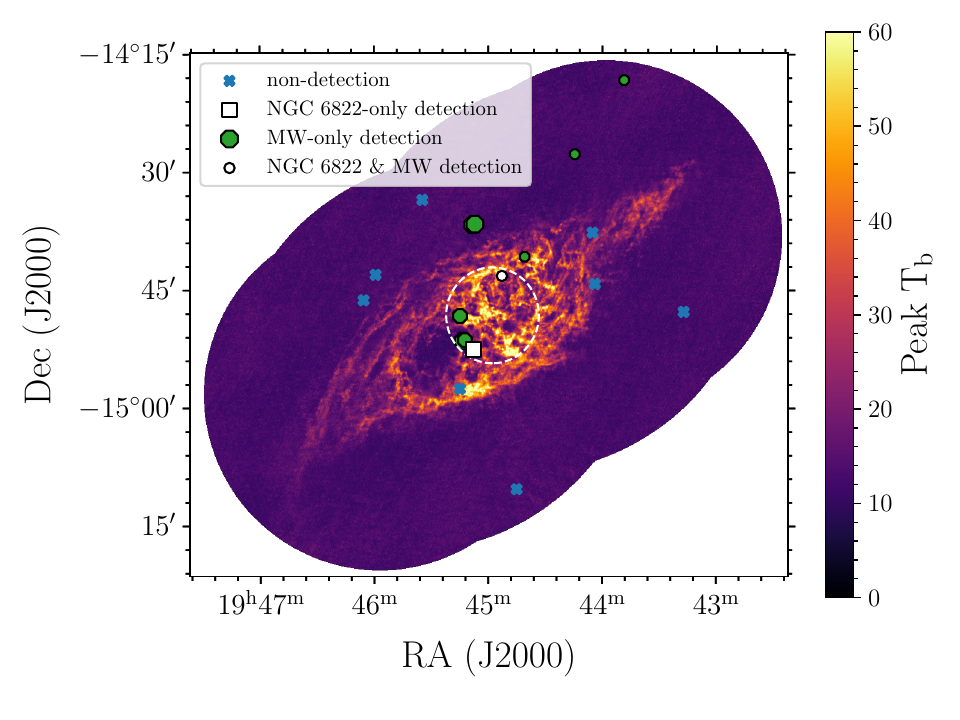}
\caption{\label{fig:ngc6822_peak} The peak \hi\ brightness temperature image of NGC 6822 from LGLBS. This image does not contain single dish observations to correct for the missing short spacings. The locations of background sources used in our search for absorption are denoted by the overlaid symbols. The different shapes classify our detections: non-detections (blue cross), absorption only at velocities corresponding to Milky Way foreground (green octagons), absorption at velocities corresponding to Milky Way foreground and NGC 6822 (white circle), and absorption only reliably detected at the velocities of NGC 6822 (white square). The symhol sizes indicate the relative flux density level of the background source. The two white symbols show the location of \hi\ absorption at the LSR velocities of NGC 6822 and are highlighted in the profiles shown in Figure~\ref{fig:abs_detections}. The dashed white circle shows the mean radius of the \hi\ column density of 6$\times10^{20}$ cm$^{-2}$.}
\end{figure*}

\begin{figure}
\centering
\includegraphics[width=\columnwidth]{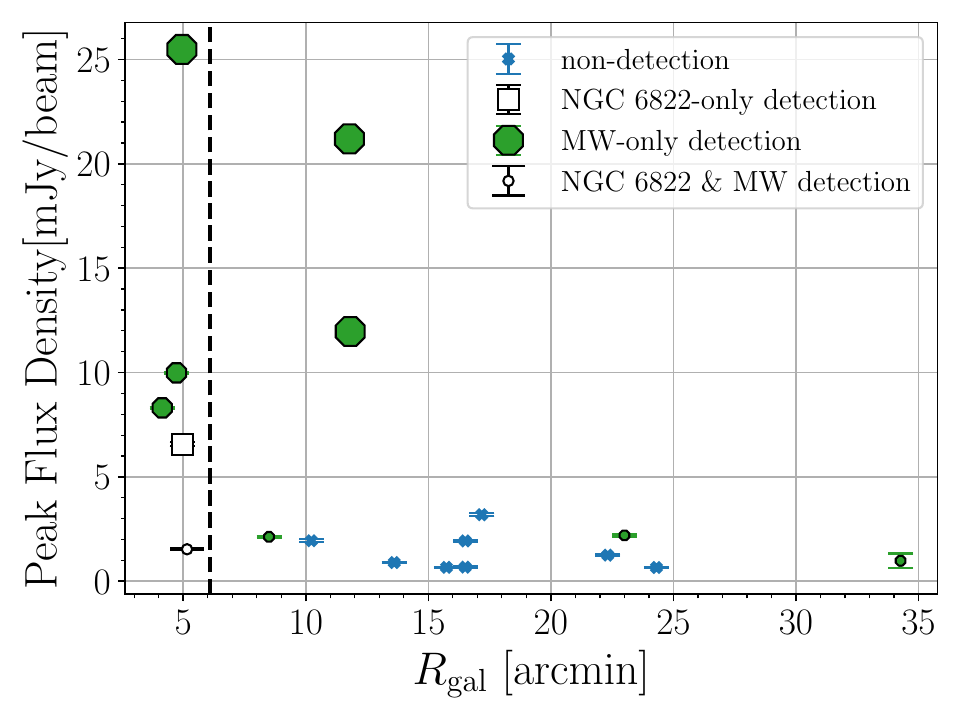}
\caption{\label{fig:ngc6822_flux_vs_Rgal} The peak continuum flux density of suitably strong background sources in our narrowband continuum image measured by the {\tt aegean} source finding algorithm as a function of projected radial distance from the optical center ($R_{\rm gal}$). The symbols and their relative sizes represent the same detection classification and flux level as described in the caption of Figure~\ref{fig:ngc6822_peak}. The vertical dashed line shows the mean radius of the \hi\ column density of 6$\times10^{20}$ cm$^{-2}$.}
\end{figure}

The peak \hi~temperature distribution for NGC 6822 is shown in Figure~\ref{fig:ngc6822_peak} with the overlaid symbols showing the location of background sources used to search for absorption from intervening cold \hi~clouds. There is a large amount of detailed small-scale structure, including \hi~shells marking the sites of energy injection into the ISM from recent supernovae and filaments where flows of denser gas converge. Figure~\ref{fig:ngc6822_flux_vs_Rgal} shows the peak flux density of each of our background sources as a function of projected distance from the optical center of NGC 6822 ($\alpha_{\rm J2000}$=19h44m57.74s, $\delta_{\rm J2000}$=-14$^\circ$48m12.363s). 72\% of our sightlines with no detections at the velocities of NGC 6822 lie outside of the mean radius of the 6$\times10^{20}$ cm$^{-2}$ threshold for star formation, while all sightlines without any absorption signal are towards sources with peak flux densities below 5 mJy/beam. Full details about the emission imaging pipeline and properties of the emission cube are discussed in N. Pingel et al. (in preparation).

\subsection{Absorption Pipeline}
\label{subsec:absorption_pipeline}


The wide-field nature of the LGLBS observations means there will be potentially hundreds of sightlines to probe for \hi\ absorption, especially for the Local Group spiral galaxies M31 and M33. By leveraging the 20000+ computing cores available through the CHTC, we designed a DAG to efficiently extract science-quality absorption and emission profiles from LGLBS data products based on the data processing pipeline developed by \citet{dempsey2022} for the GASKAP-\hi~survey. We use only measurement sets from the extended A and B configurations to ensure that we filter out large-scale emission structure that may affect the stability of our continuum level. Below we outline a high-level description for the major components of the DAG with the associated CASA task names in a monospace font and note that the code and documentation are available as part of a public {\tt GitHub} repository\footnote{\url{https://github.com/nipingel/LGLBS_Absorption}}. 
\begin{enumerate}
    \item \textit{Pre-processing}:~split out \hi\ spectral window {\tt split()}, resample all visibilities on the same spectral axis and spectral reference frame {\tt mstransform()}, combine into a single measurement set {\tt concat()}, and re-weight visibilities based on the rms measured in signal-free channels {\tt statwt()}.
    \item \textit{Narrow-band continuum image}: We generate a multifrequency synthesis image of the source across a 4 MHz bandwidth centered on 1420.6946 MHz in the LSR reference frame with channels corresponding to the source and foreground Milky Way emission flagged out and use 20000 Hogbom clean iterations. We refer to this as the narrowband continuum image and obtain a restoring beam size for the image of 3.26$''\times2.36''$.
    \item \textit{Source catalog}: To identify continuum sources, we run  {\tt aegean}\footnote{\url{https://github.com/PaulHancock/Aegean}}, an automated source finding package \citep{hancock2012, hancock2018} for radio astronomy sources. This software is designed for use on wide-field images where many parameters may vary across its extent, in addition to fitting models of source structure that are spatially correlated. We adopt the default parameters and produce a list of background sources with source positions, source semi/major axis information, and flux density.
    \item \textit{Absorption-line extraction}: After parsing the source catalog to select suitably strong background sources (see below), we produce a \hi\ spectral line sub-cube (we will refer to these as ``cubelets''), spanning a field of 66$''\times66''$ centered on the source. When constructing this cubelet, we include only baselines with lengths above 1 km --- corresponding to projected physical scales $\leq$ 100 pc at the distance to NGC~6822 --- to ensure as much extended emission is resolved out as possible, allowing us to focus on the spectrum associated with the compact background source. We experimented with cutoffs up to 1.5 km and found consistent results. Each cubelet was cleaned for 1000 Hogbom iterations with {\tt tclean()}, has 0.75$''\times0.75''$ pixels and 700 spectral channels spanning a total of 300 \kms, and was weighted naturally, resulting in a Gaussian restoring beam size of 5.5$''\times3.6''$. We extract \hi\ absorption spectra in the direction of each radio continuum source by analyzing each cubelet. We define a radial pixel grid centered on the position of each source. Only spatial pixels that fall within the ellipse that defines the extent of the source as determined by the source finding algorithm are used to construct the profile. For the relevant pixels, we measure the mean brightness temperature along each line-of-sight in line-free regions of the cubelet ($-$200~\kms$\leq v_{\rm LSRK}\leq-150$~\kms~and $+$35~\kms$\leq v_{\rm LSRK}\leq+100$~\kms). For each spectral channel, we sum the pixel values while weighting their contribution to the final spectra by the square of the line-of-sight mean brightness temperature \citep{dickey1992}. Lastly, we place the absorption spectra in units of optical depth ($e^{-\tau}$) by dividing by its mean value measured across line-free velocities. 
    \item \textit{Source selection}: Initially, we identify 151 individual radio components in the narrow-band continuum image with peak flux densities ranging from 0.2 mJy/beam to 54 mJy/beam. The rms noise level measured in the image plane of emission-free channels in the cubelets is 0.55 mJy/beam. To achieve a 3$\sigma$ detection in optical depth in units of $e^{-\tau}$, we require the 1$\sigma$ noise in the optical depth spectrum to be less than 0.33 (see Equation 1 below and note that, ideally, a noise-free and signal-free optical depth spectrum is normalized to 1.0). Given the measured noise level in the cubelets, sources stronger than 0.55 mJy/beam/0.33=1.67 mJy/beam are therefore required. After accounting for the differences in beam sizes between the narrowband continuum image (3.26$''\times2.36''$) and cubelets (5.5$''\times3.6''$), we parse the initial source catalog to focus on the cubelets from sources with cataloged peak flux density levels greater than 0.65 mJy/beam, reducing our total radio sources against which we probe for \hi\ absorption from 151 to 18.
    \item \textit{Emission-line extraction}: The calculation of important physical properties, such as the spin temperature, requires knowledge of the \hi\ emission near the absorption detection. For each suitable background source, we first extract a 30$'$ in diameter subregion around each source from the continuum-subtracted emission cube (N. Pingel et al., in preparation). We account for the missing short-spacings by combining our VLA data with single dish observations from the Green Bank Telescope (project ID: AGBT13B\_169). Detail of the feathering process are discussed in detail in E Koch et al. (in preparation). We construct the final \hi\ emission spectrum by measuring the mean brightness temperature within an annulus of 14$''$ (20 pixels, $\approx$ 2 beam widths) centered on each source, with the central 7$''$ (10 pixels, $\approx$ 1 beam width) excluded from the calculation.
\end{enumerate}

\subsection{Noise Estimates}\label{subsec:noise}We estimate the noise in the absorption spectra by following the method outlined in \citet{roy2013}. We first measure the standard deviation in the signal-free velocities ($\sigma_{\rm cont}$). Even though we utilize only baselines above 1 km, interferometers are not perfect high-pass filters, meaning some extended emission will be detected and raise the system temperature in spectral channels containing significant emission. We model this increase in system temperature by measuring the average brightness temperature spectrum ($T_{\rm em}(v)$) in each of the extracted and feathered emission sub-cubes that span the 30$'$ extent of the VLA primary beam at L-Band. The 1$\sigma$ optical depth noise spectrum in units of $e^{-\tau}$ (the grey band in the top panels of figure~\ref{fig:abs_detections}) is calculated as:
\begin{equation}\label{eq:optical_depth_noise}
\sigma_\tau(v)=\sigma_{\rm cont}(v)\frac{T_{\rm sys}+\eta_{\rm ant}T_{\rm em}(v)}{T_{\rm sys}}, 
\end{equation}
where the system temperature $T_{\rm sys}=25$ K and antenna efficiency $\eta_{\rm ant}=0.35$. The final rms noise in units of $e^{-\tau}$ ($\sigma_{\tau}$) for a given absorption spectrum is taken to be the rms of the noise spectrum in signal-free spectral channels.

\section{HI absorption in NGC 6822}\label{sec:results}

Figure~\ref{fig:ngc6822_peak} shows the position of 18 radio sources from the 4 MHz narrowband continuum image with peak flux density values ($S_{\rm peak}$) ranging from 0.66 mJy/beam to 54 mJy/beam, overplotted on the \hi\ peak brightness temperature image. Eight of these sources probe the central region of NGC6822, with $T_b$ of the nearby \hi\ emission peaking at 90 K. We detected clear \hi\ absorption only in the direction of two sources at velocities consistent with NGC~6822 shown as white symbols in Figure~\ref{fig:ngc6822_peak}. Our typical 1$\sigma$ optical depth sensitivity in units of $e^{-\tau}$, computed by taking the median rms across signal-free channels across these 18 sightlines, is 0.05. This value is consistent with the best optical depth noise achieved by \citet{dempsey2022}, which probes \hi\ absorption in the SMC, and roughly a factor of two more sensitive than the survey of \citet{dickey1993} towards M31 and M33. This sensitivity is also similar to that of many Milky Way \hi\ absorption surveys, e.g. \cite{Heiles2003}, and matches the lowest noise levels in the sample presented by \citet{dickey2022}, but is about an order of magnitude less sensitive than the highly sensitive, high-latitude Milky Way \hi\ absorption survey by \citet{murray2018}. We note that, because optical depth sensitivity depends on the strength of the background source used to probe for absorption, our sensitivities ultimately depend on the properties of the unique set of background sources within our field of view.



\subsection{Detections}\label{subsec:detections}
\begin{figure*}[t]
\includegraphics[width=0.5\textwidth]{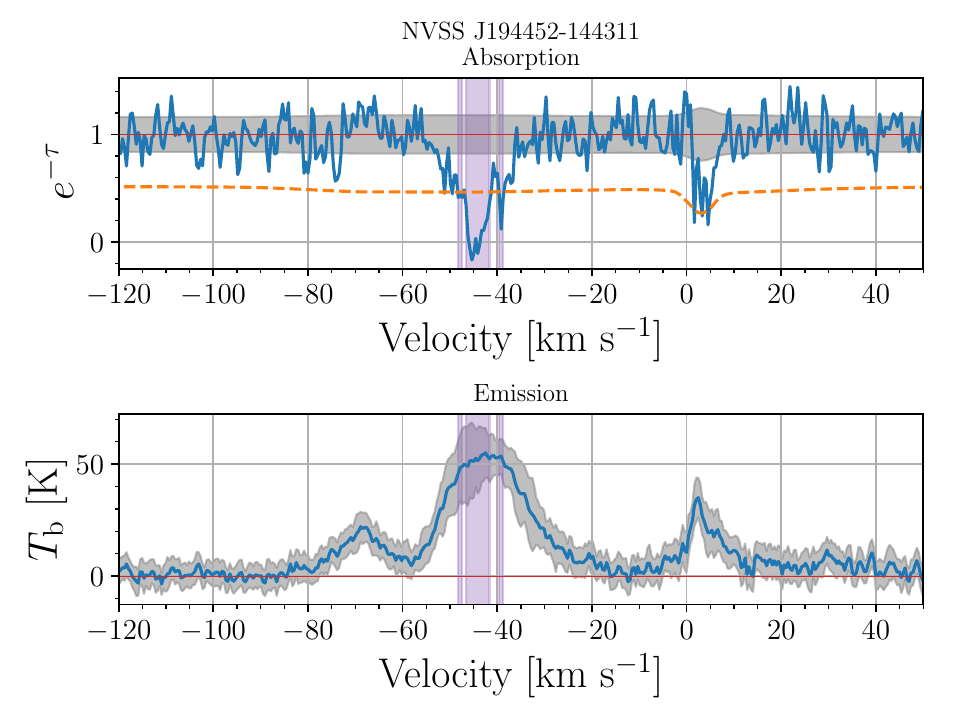}
\includegraphics[width=0.5\textwidth]{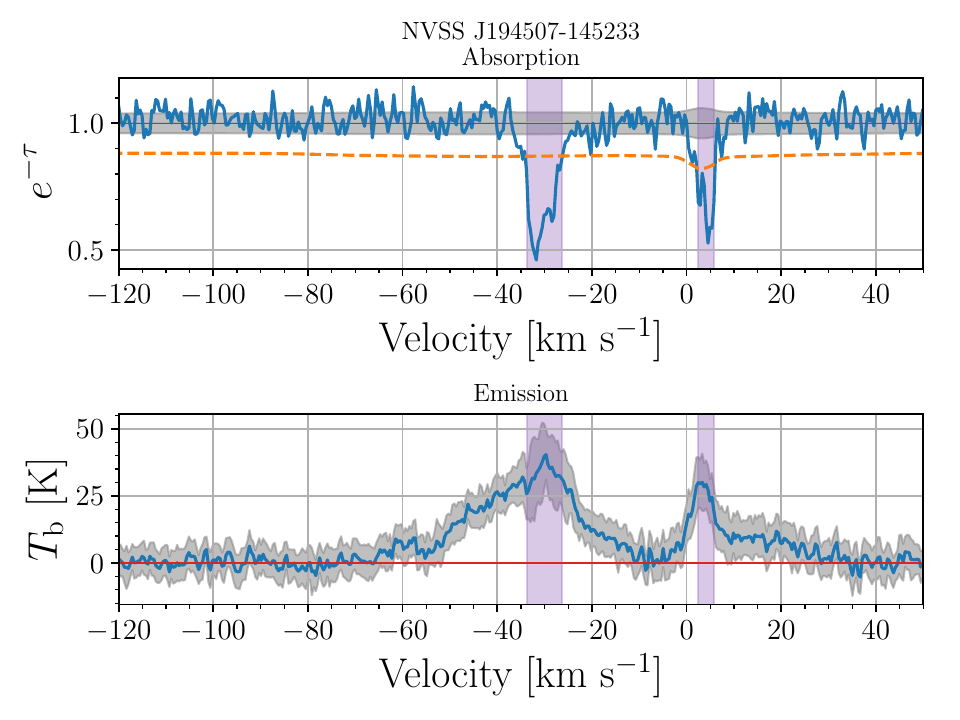}
\caption{\label{fig:abs_detections} The absorption (top) and associated emission (bottom) spectra constructed using the methods outlined in Section~\ref{subsec:absorption_pipeline} for our sightlines with detections of \hi\ absorption at NGC 6822 velocities ($<-10$ \kms). The red horizontal line in the absorption spectra highlights the $e^{-\tau}$ level of 1.0; the grey shaded region and orange dashed line respectively show the 1$\sigma$ (in both emission and absorption) and 3$\sigma$ level (absorption-only); and the purple shaded regions highlight the spectral channels with \hi\ absorption exceeding this 3$\sigma$ level.} 
\end{figure*}

The absorption and associated emission spectra for our two detections are shown in Figure~\ref{fig:abs_detections}. In both cases, the background continuum sources have been previously cataloged in the NRAO VLA Sky Survey (NVSS; \citealt{condon1998}) with names NVSS J194452-144311 and NVSS J194507-145233 and total fluxes 19.4 mJy and 33.8 mJy for the northern and southern sources, respectively. We measure total fluxes of 5.2 mJy and 19.3 mJy, with the differences largely due to the superior rms confusion flux density of LGLBS.

The two detections have peak optical depth (in units of $\tau$) of 0.7 to $\sim3$, which again agrees with previous work within the SMC and other Local Group galaxies \citep{dickey1993, dempsey2022}. In both directions, we see indications of \hi\ absorption at foreground Milky Way velocities, with the more pronounced signal towards the southern source. The \hi\ emission profiles for both sources, especially for the second source, are broad, reflecting emission from a mixture of cold and warm gas that remains spatially unresolved or crowded in velocity along these sightlines. Clearly, constraints on both the cold and warm phases require joint modeling. 

The total optical depth-corrected \hi\ column density, where 
\begin{equation}\label{eq:HIcolumndensity}
N_{\hi\,\rm thick}=1.823\times10^{18}\int \frac{T_{b}\tau}{1-e^{-\tau}}dv,
\end{equation}
for the northern and southern detections measured over $-100$ \kms\ to $-5$ \kms\ are respectively 3.2$\times10^{21}$ cm$^{-2}$ and 1.8$\times10^{21}$ cm$^{-2}$, exceeding the commonly adopted star-formation threshold of $6 \times 10^{20}$ cm$^{-2}$ to $1\times 10^{21}$ cm$^{-2}$ (e.g., \citealt{krumholz2009a, wolfire2010, sternberg2014}) needed to shield molecular gas from photodissociation depending on the geometry (i.e., beamed or isotropic) of the background radiation fields. The ratio of optical-depth corrected \hi\ column density to that derived assuming the emission is optically thin, where $N_{\hi\,\rm thin}$=1.82$\times$10$^{18}\int T_{b}dv$, is respectively 1.2 and 1.13 for the northern and southern detections. The northern and southern detections respectively have optical depth rms values of 0.11 and 0.03, translating to a $3\sigma$ $N_{\hi\,\rm thick}$ detection threshold of 6.0$\times10^{19}$ cm$^{-2}$ and 5.9$\times10^{19}$ cm$^{-2}$ over a 5.0 \kms\ line width. The NVSS names of the source, Right Ascension (RA) and Declination in J2000 coordinates, total flux, and $\sigma_{\tau}$ are summarized in the first 5 columns of Table~\ref{tab:properties}. 

\subsection{Non-Detections}\label{subsec:non-detections}

\begin{figure}
\centering
\includegraphics[width=\columnwidth]{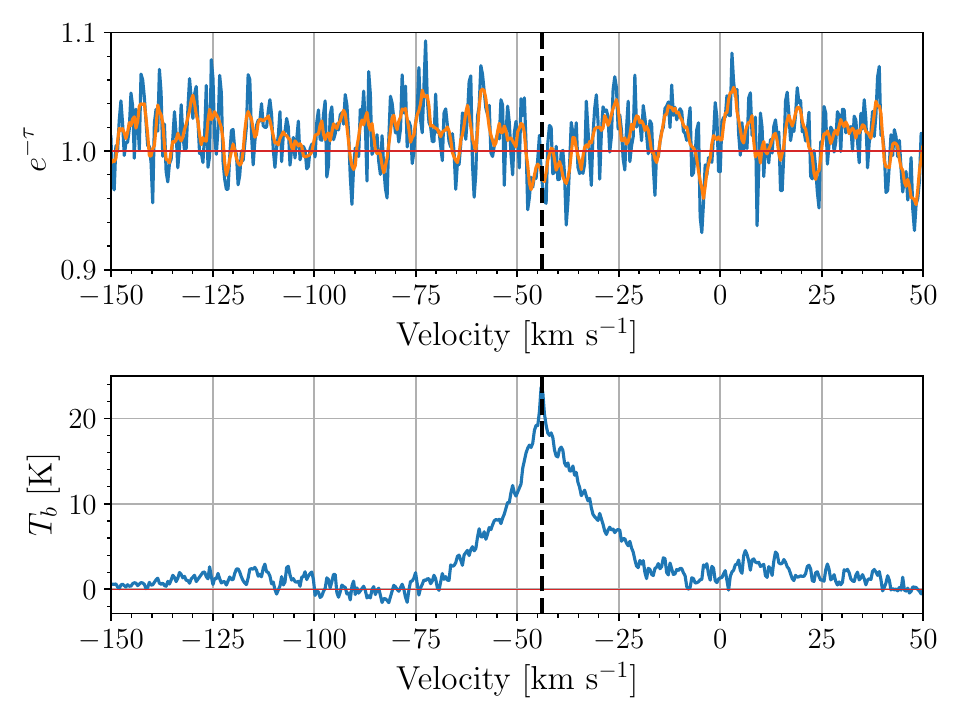}
\caption{\label{fig:stacking_analysis} The stacked absorption spectrum (top) and emission spectrum (bottom) for 13 sightlines with no detections of \hi\ absorption at velocities associated with NGC~6822. The applied velocity shift is the offset between the $-$44 \kms\ systemic velocity of NGC 6822 (vertical dashed line) and the velocity at peak intensity of the \hi\ emission between $-$150 \kms\ to $+50$ \kms in the LSRK reference frame. We fit and subtract Gaussian functions to the emission and absorption features between $\pm$5 \kms\ in eight emission/absorption pairs to avoid introducing bias from Milky Way foreground gas. The orange profile in the top panel shows the same stacked absorption profile after applying a 5-pixel wide boxcar smoothing.}
\end{figure}

A total of 16 of the 18 sightlines we investigated show no indication of \hi\ absorption near the $-$44 \kms\ systemic velocity of NGC 6822. For all sightlines without detections, $N_{\hi,\rm thin}$ ranges from 2.5$\times10^{20}$ cm$^{-2}$ to 1.5$\times10^{21}$ cm$^{-2}$ with only four non-detection sightlines exceeding the star-formation threshold and intersecting the primary \hi\ disk of NGC 6822. 

We search for extremely weak absorption signals at the expected velocities of NGC 6822 by employing the spectral stacking techniques used by \citet{murray2014}. Stacking spectra requires determining an appropriate velocity shift to align potential signals around the local systemic velocity of NGC~6822 and so remove the effects of galactic rotation and other large-scale flows. Then the aligned spectra can be coherently averaged. Generally, the velocity shift is determined by computing the first velocity moment of each \hi\ emission profile. But there is substantial overlap in the radial velocity components between \hi\ from NGC 6822 and \hi\ from the Milky Way, so overlapping signals must be carefully taken into account to avoid misidentifying a stacked signal that arises strictly from foreground gas. Fortunately, all of our sightlines with detected \hi\ emission are similar to those in the bottom panels of Figure~\ref{fig:abs_detections} such that peaks associated with NGC 6822 and the Milky Way are separated in velocity by 10 \kms or more. We remove the Milky Way \hi\ emission components by fitting and subtracting a single Gaussian function, restricted to having a mean within $\pm$5 \kms in the LSRK velocity frame. Similarly, we fit and subtract either one or two Gaussian functions to the Milky Way \hi\ absorption features within the same velocity range. We determine the preferred model by comparing their respective reduced chi-squares.

We need to determine the optimal velocity shift after removing our models of the Milky Way emission and absorption components. The stacking analysis in \citet{koch2019} reveals the \hi\ and CO ($2-1$) velocity at peak intensity to be highly correlated in M33. To maximize the possibility of extracting a stacked \hi\ absorption signal from CNM, we determine the velocity shift to be $\Delta v=-v_{\rm peak}-v_{\rm sys}$, where $v_{\rm peak}$ is the velocity of the peak \hi\ intensity in and the systemic velocity of $-$44 \kms LSRK. We visually inspect the \hi\ emission spectra in each of the 16 non-detection sightlines and remove an additional three emission/absorption spectra from the stacking analysis due to lack of emission signal, as they lie considerably outside of the main body of NGC 6822. In total, we therefore stack 13 emission/absorption spectra with eight instances where models for the foreground Milky Way emission and absorption components are subtracted out. As in \citet{murray2014}, we maximize signal-to-noise by applying a weight of $W=1/\sigma_{\tau}$, where $\sigma_{\tau}$ is taken as the rms of $e^{-\tau}$ measured in signal-free velocities of $-$200 \kms\ to $-150$ \kms and 35 \kms\ to 50 \kms. We multiply each absorption and emission spectra by these associated weights, sum over all $n=13$ profiles, and normalize by $\Sigma_{n}W_{n}$. 

The resulting stacked absorption and emission spectra are shown respectively in the top and bottom panels of Figure~\ref{fig:stacking_analysis}. While we increase our sensitivity (in units of $1-e^{-\tau}$) by a factor of $\sim$4 relative to the median rms noise calculated in the same velocity channels in the individual non-detection velocity profiles (0.013 as opposed to 0.050), there is no discernible signal revealed by the stacking at the systemic velocity of NGC 6822. This is still true after smoothing the stacked absorption spectrum by a 5-pixel wide boxcar function to enhance the signal-to-noise. In units of $\tau$, we are able to place a $3\sigma$ upper limit on the \hi\ absorption in these spectra at a level of $\tau\leq0.04$.


While no underlying weak absorption is detected from the non-detections in NGC 6822 in this study, our stacking analysis demonstrates the quality stability and flatness of the spectra extracted using the LGLBS absorption pipeline. We will build on this stacking analysis by including spectra from non-detections in other LGLBS sources to maximize the possibility of revealing low-level absorption. 

\section{Properties of Individual cold \hi\ structures}\label{sec:indv_CNM}

\subsection{Gaussian Decomposition}\label{subsec:gauss_decomp}


We decompose emission and absorption spectra to estimate physical properties of the individual clouds along the line of sight. The method we use here originates from \cite{Heiles2003}.

We assume \hi\ clouds are multi-phase consisting of CNM and the WNM. The optical depth (absorption spectrum) only comes from the CNM, while the brightness temperature (emission spectrum) is produced by the combined contribution from the  CNM and the WNM. 
We first fit the optical depth spectrum $\tau(v)$ with a set of $N$ Gaussian-like functions, corresponding to $N$ CNM components:
\begin{equation}
\tau(v)=\sum_0^{N-1} \tau_{0, n} e^{-\left[2\left(v-v_{0, n}\right) / \delta v_n\right]^2}
\end{equation}
where $N$ is the minimum number of components necessary to make the residuals of the fit smaller or comparable to the estimated noise level of $\tau(v)$. $\tau_{0, n}$ is the peak optical depth, $v_{0, n}$ is the central velocity, and $\delta v_n$ is the $1 / e$ width of component $n$. 

The total expected brightness temperature including contributions from both the CNM and the WNM components is:
\begin{equation}
T_{\exp }(v)=T_{B, \mathrm{CNM}}(v)+T_{B, \mathrm{WNM}}(v),
\end{equation}
where
$T_{B, \mathrm{CNM}}(v)$ is the \hi\ emission from $N$ CNM components which can be represented as:
\begin{equation}
T_{B, \mathrm{CNM}}(v)=\sum_0^{N-1} T_{s, n}\left(1-e^{-\tau_n(v)}\right) e^{-\sum_0^{M-1} \tau_m(v)}
\end{equation}
where $\tau_m(v)$ represents each of the $M$ CNM clouds that lie in front of cloud $n$ 
and $T_{s,n}$ is the spin temperature of cloud $n$.
For multiple absorption components, we fit all possible orders along the line of sight and choose the one that yields the smallest residuals. The spin temperature $T_{s,n}$ is generated from this equation. 

For the WNM part in the emission profile, we use $K$ Gaussian functions to represent the original unabsorbed emission from the WNM. We account for the CNM absorption by assuming that a fraction $F_k$ of the WNM remains unabsorbed in front of all CNM components, with the remainder located behind them:
\begin{equation}
T_{B, \mathrm{WNM}}(v)=\sum_0^{K-1}\left[F_k+\left(1-F_k\right) e^{-\tau(v)}\right] \times T_{0, k} e^{-\left[\left(v-v_{0, k}\right) / \delta v_k\right]^2}
\end{equation}
where $T_{0, k}$, $v_{0, k}$ and $\delta v_k$ represent the Gaussian fitting parameters (peak in units of brightness temperature, central velocity, and 1/e linewidth) of the original unabsorbed $k$-th WNM component. We examine three specific values for the fraction $F_k: (0.0, 0.5, 1.0)$. The two extremes, where $F_k=0$ and $F_k=1$, correspond to situations where all WNM lies behind the CNM, resulting in the WNM being totally absorbed, and all WNM lies in front of CNM, resulting in a complete lack of absorption, respectively. Essentially this places lower and upper limits on $T_{B,WNM}$. 

There are a total of $N!$ possible orderings for the N CNM components. With $k$ WNM components, and three possible values for $F_k$, there are a total of $3^k$ combinations for the WNM. Thus, the total number of possible orderings is given by $[N!3^k]$. The final best fit is selected with the least residuals, and the final spin temperatures ($T_{\rm s}$) are calculated by a weighted average over all trials (cf. Equations (21a) and (21b) of \cite{Heiles2003}).

\begin{figure*}
\centering
\includegraphics[width=0.9\textwidth]{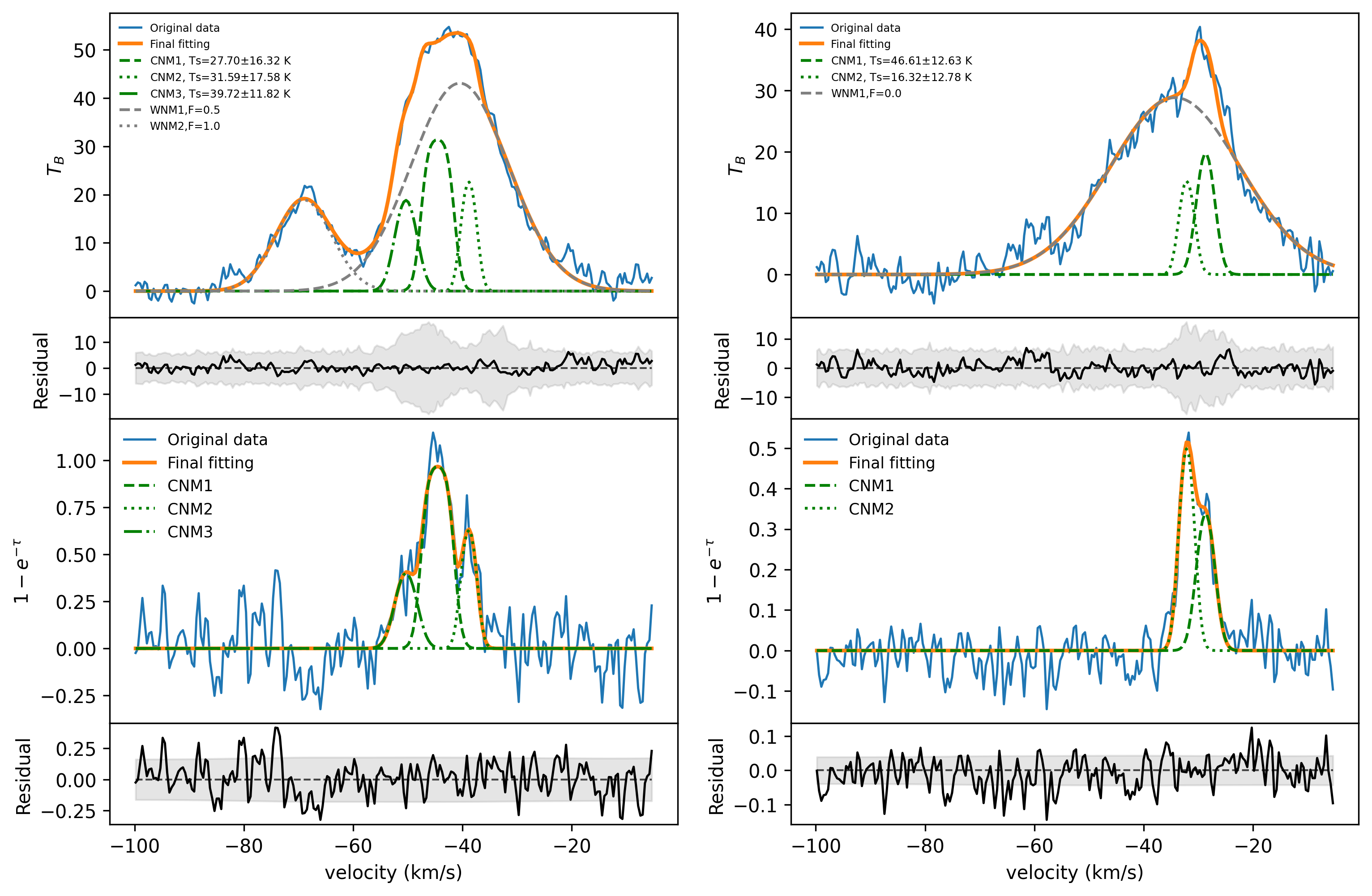}
\caption{\label{fig:gaussian_decomp}Gaussian decomposition for two absorption-detected sources. The left is source NVSS J194452-144311 and the right source is NVSS J194507-145233. The top two panels show the \hi\ brightness temperature and its corresponding fitting residuals. The bottom two panels present the \hi\ optical depth and its fitting residuals. The top panel specifies the weighted mean spin temperature of the CNM. The `F' value following each `WNM' label quantifies the fraction of the WNM absorbed by the CNM, as stated in the text. The gray shaded area in the residual indicates the 1$\sigma$ error of the emission/absorption spectra.}
\end{figure*}

Our best fitting results are shown in Figure~\ref{fig:gaussian_decomp}, with the corresponding parameters detailed in Table~\ref{tab:properties}. It is important to note that part of the absorption spectra for NVSS J194452-144311 is saturated. Given that our fitting procedure targets the $\tau$, the saturation will inevitably become an issue when applying the Gaussian fit to $\tau$. To address the potential infinite values when converting $1-e^{-\tau}$ to $\tau$, we set $1-e^{-\tau}=0.96$ for those velocity channels that are saturated. This is the closet value to 1.0 that both avoids infinite values when converting to units of $\tau$ and allows our fit to converge. To help with fitting, we slightly increase the signal-to-noise by smoothing the absorption spectrum towards NVSS J194452-144311 using \texttt{astropy.convolution.Gaussian1DKernel} with a standard deviation of 0.5 pixels. This slightly decreases our velocity resolution from 0.4 \kms\ to 0.5 \kms.

For each component identified, key fitting parameters listed in Table~\ref{tab:properties} include peak velocity ($v_{\text{peak}}$), the full width half maximum (FWHM; $\Delta v_{\rm FWHM}$), the spin temperature from the fit ($T_{\text{s}}$) for CNM, and peak optical depth ($\tau_{\text{peak}}$) for CNM or peak brightness temperature ($T_{\text{b,\rm peak}}$) for the WNM. We also calculate the maximum kinetic temperature ($T_{k,\rm max}=21.86\Delta v_{\rm FWHM}^2$) based on the line widths and \hi\ column density ($N_{\hi}$). For the CNM, the \hi\ column density is given by $N_{\text{\hi,CNM}}=1.823 \times 10^{18} \times  T_s \int \tau d v \mathrm{~cm}^{-2} $, and for the WNM, it is $N_{\text{\hi,WNM}}=1.823 \times 10^{18}  \int T_B d v \mathrm{~cm}^{-2} $. We derive the CNM fraction as $f_{\text{CNM}} = N_{\text{HI,CNM}}/ (N_{\text{\hi,CNM}} + N_{\text{\hi,WNM}}$). We do not constrain or report $T_{s}$ for the fitted WNM components. For comparison with other literature, we also include the density-weighted mean spin temperature \citep{dickey2000}:
$\left\langle T_{\rm s}\right\rangle=\frac{\int T_{\mathrm{B}}(\mathrm{v}) d \mathrm{v}}{\int 1-e^{-\tau(\mathrm{v})} d \mathrm{v}}$.

To assess the uncertainties associated with $\left\langle T_{\rm s}\right\rangle$, we employ a Monte Carlo method. Specifically, we generate 1000 samples for both the optical depth and the brightness temperature, drawing randomly from a normal distribution where the actual spectrum value serves as the mean and the rms noise defines the standard deviation. We then execute all calculations for each sample, adopting the mean of the outcomes as the final value for $\left\langle T_{\rm s}\right\rangle$ and the standard deviation of these results to represent the $1\sigma$ uncertainty.

\begin{table*}[ht]
\centering
\caption{Source and Fitting Properties}
\label{tab:properties}
\resizebox{\textwidth}{!}{
\begin{tabular}{ccccccccccccccc}
\hline
Source  & RA  & Dec & $S_{\rm peak}$ & $\sigma_{\tau}$  &$f_{\text{CNM}}$& $\left\langle T_{s}\right\rangle$ & Phase \& Component Number & $T_{\text{s}}$ & $T_{k,\rm max}$ & $\tau_{\text{peak}}$ or $T_{b,\rm peak}$ [K] & $v_{\text{peak}}$ & $\Delta v_{\rm FWHM}$  & $N_{\hi}$ \\
& [hms] & [dms] & [mJy]  &  &  & [K] & & [K] & [K] & & [\kms] & [\kms] & 10$^{20}\text{cm}^{-2}$ \\ \hline
NVSS J194452-144311 & 19h44m52.78s &-14d43m11.88s & 19.4 & 0.11 & 0.330 & 414$\pm$54 & CNM1 & 27.70$\pm$16.32 & 346.93 &3.44 &$-$44.54 &3.98 &7.37\\ 
& & & & & & & CNM2 &  31.59$\pm$17.58 & 178.23 & 1.00 &$-$38.84 &2.86 & 1.74 \\
& & & & & & & CNM3 & 39.71$\pm$11.82 & 416.02 & 0.51 &$-$50.33 &4.36 & 1.72 \\
& & & & & & & WNM1 &  & 9611.40& 43.09 &$-$40.49 & 20.97 &  17.53  \\
& & & & & & & WNM2 &  & 3245.08& 18.91 &$-$69.05 & 12.18 &  4.47 \\
\hline
NVSS J194507-145233 & 19h45m7.70s& -14d52m32.19s & 33.8 & 0.03 & 0.115 & 138$\pm$15 & CNM1 & 46.61$\pm$12.63 & 294.78 & 0.41 &$-$28.66 &3.67 &1.37\\ 
& & & & & & & CNM2 & 16.31$\pm$12.78 & 194.74 & 0.69 &$-$32.12 &2.98 & 0.65 \\
& & & & & & & WNM1 & & 17145.83& 28.85 &$-$34.25 & 28.00 &  15.56  \\
 \hline
\end{tabular}
}
\end{table*}

\subsection{Spin temperature and CNM fraction}

The detected cold \hi\ has $\Delta v_{\rm FWHM} \sim$ 3-4 km/s and $\tau_{\rm peak} \sim$ 0.4 to 3.4 --- both similar to what is found in the SMC \citep{jameson2019, dempsey2022}. The implied $T_{k, \rm max}$ ranges between 200 K and 400 K. The estimates for $T_{s}$ fall within 20 K and 50 K, which agrees well with what was found for the SMC (\citealt{dickey2000, jameson2019, dempsey2022}, H. Chen et al. in preparation). We note that in several cases $T_s$ has large uncertainties due to the fact that emission features are not well-resolved relative to the pencil-beam measurements of the absorption. The estimated $N_{\hi,\rm CNM}$ for the detected absorption features is 0.6--2 $\times 10^{20}$ cm$^{-2}$, with only one component having $7\times 10^{20}$ cm$^{-2}$. The estimated CNM fraction in two directions is $0.33$ and $0.12$ which is slightly higher than what has been found for the SMC (typically $<15$\%). In terms of properties of the detected cold \hi\, we find many similarities between NGC 6822 and the SMC. 

\section{Association with Other ISM Tracers}\label{sec:multiwavelength_comparison}
\begin{figure*}[t]
\includegraphics[width=0.5\textwidth]{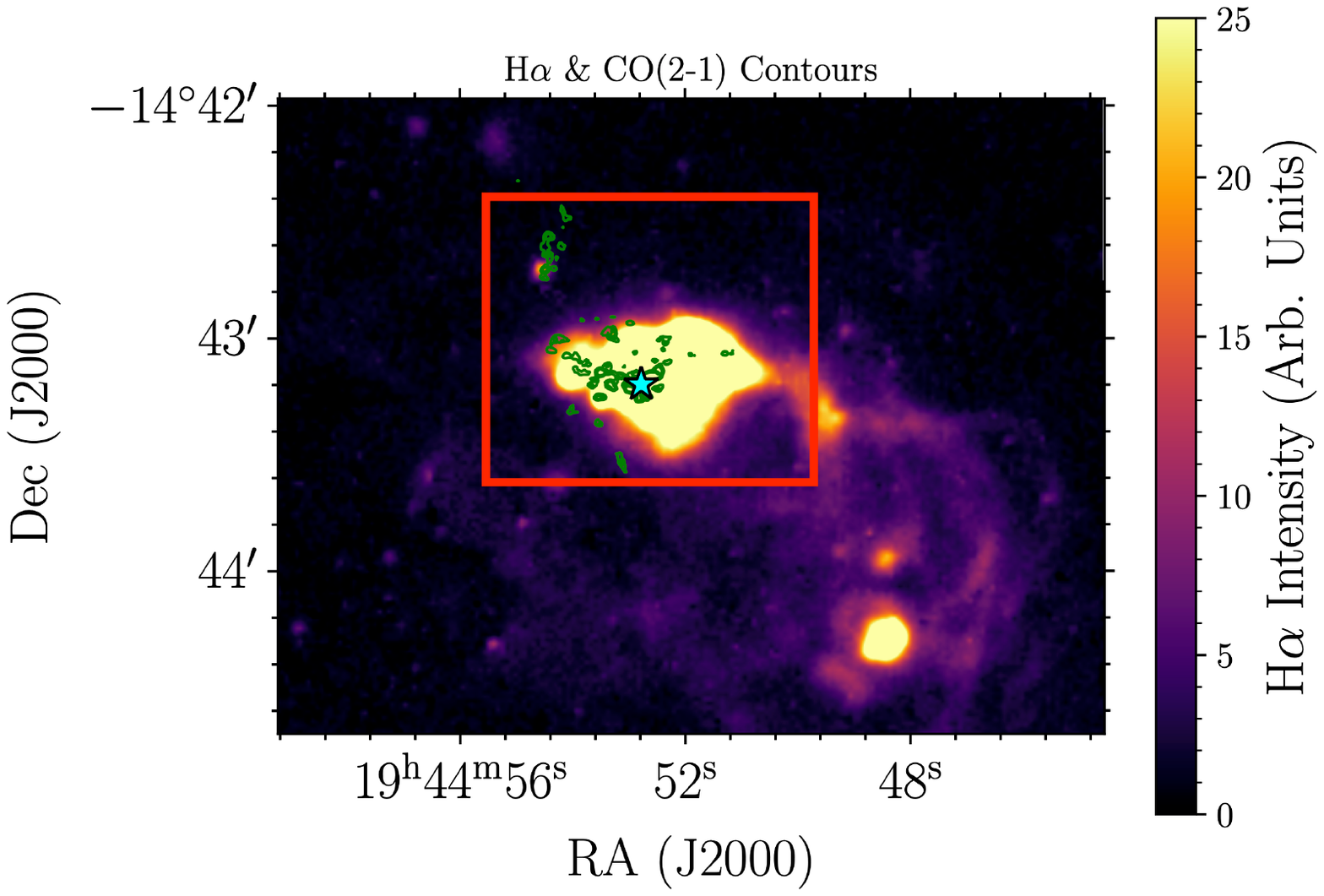}
\includegraphics[width=0.45\textwidth]{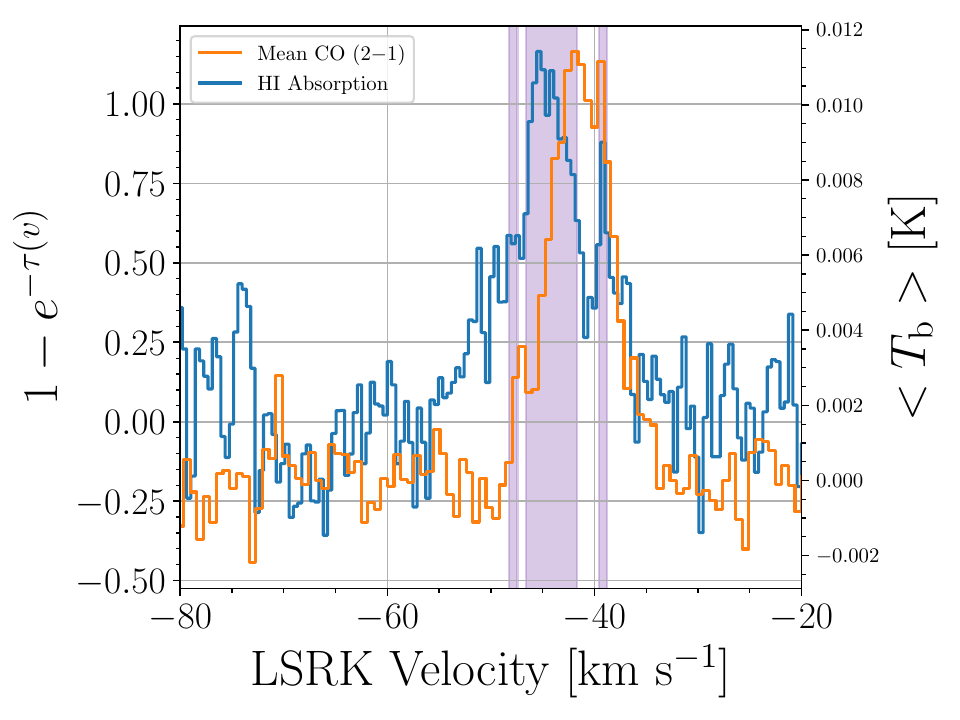}
\caption{\label{fig:halpha_CO_HI_abs}Left: Map of the H$\alpha$ intensity from \citet{hunter2004} with the location of the northern \hi\ absorption (cyan star) and contours of CO (2$-$1) observations from \citet{schruba2017} at levels of 5, 10, and 15 K \kms overlaid. Right: the \hi\ absorption spectra (blue) compared with a mean brightness temperature spectrum of CO (2$-$1) (orange) measured within the red box overplotted on the H$\alpha$ map. The purple shaded regions represent spectral channels with absorption signal exceeding the 3$\sigma$ level. The left y-axis shows units of optical depth, while the right y-axis shows the mean brightness temperature of the CO.} 
\end{figure*}

While it is predominately molecular hydrogen (\htwo) that fuels star formation via collapse, the abundance of CNM within active star-forming regions plays a role in setting the local SFR and star formation efficiency by shielding the star-forming molecular phase from photodissociation caused by penetrating far-ultraviolet (FUV) radiation. The left panel of Figure~\ref{fig:halpha_CO_HI_abs} shows that our sightline towards NVSS 194452-144311 intercepts a substantial star forming region, containing over 100 OB-type stars and a significant fraction (22\%) of the total H$\alpha$ luminosity from NGC 6822 \citep{hunter2004}. The line of sight is also spatially coincident with CO (2$-$1) emission tracing molecular gas \citep{schruba2017}. If the CNM is associated with and shielding the molecular gas, we would expect the \hi\ absorption and CO to share similar spectral signatures. The average CO spectrum, taken over the red rectangular region marking the field-of-view (FoV) of the CO observations, overlaps the \hi\ absorption profile well, though there is a $\lesssim$5 \kms~offset between their respective peaks. The slightly broader \hi\ absorption profile fully encompasses the CO spectral signature, indicating that CO-bright molecular gas is associated with the CNM probed by this sightline. This is in agreement with the findings of \citet{park2022}, which shows that CO-bright molecular gas preferentially forms in high \hi\ column density environments within the solar neighborhood. On the other hand, \citet{busch2021} and \citet{rybarczyk2022} respectively show that the broad wings in OH emission profiles and HCO$^+$ absorption profiles trace \hi\ signal but not CO, indicating that the gas in the broader, fainter wings may trace CO-dark molecular gas.  Regardless, we demonstrate a correlation of CNM with the molecular phase in a low-metallicity environment, where the lack of prevalent cooling through metal fine-structure line emission could suppress their formation \citep{glover2011}.

\subsection{HI self-absorption}

\begin{figure*}
\includegraphics[width=\textwidth]{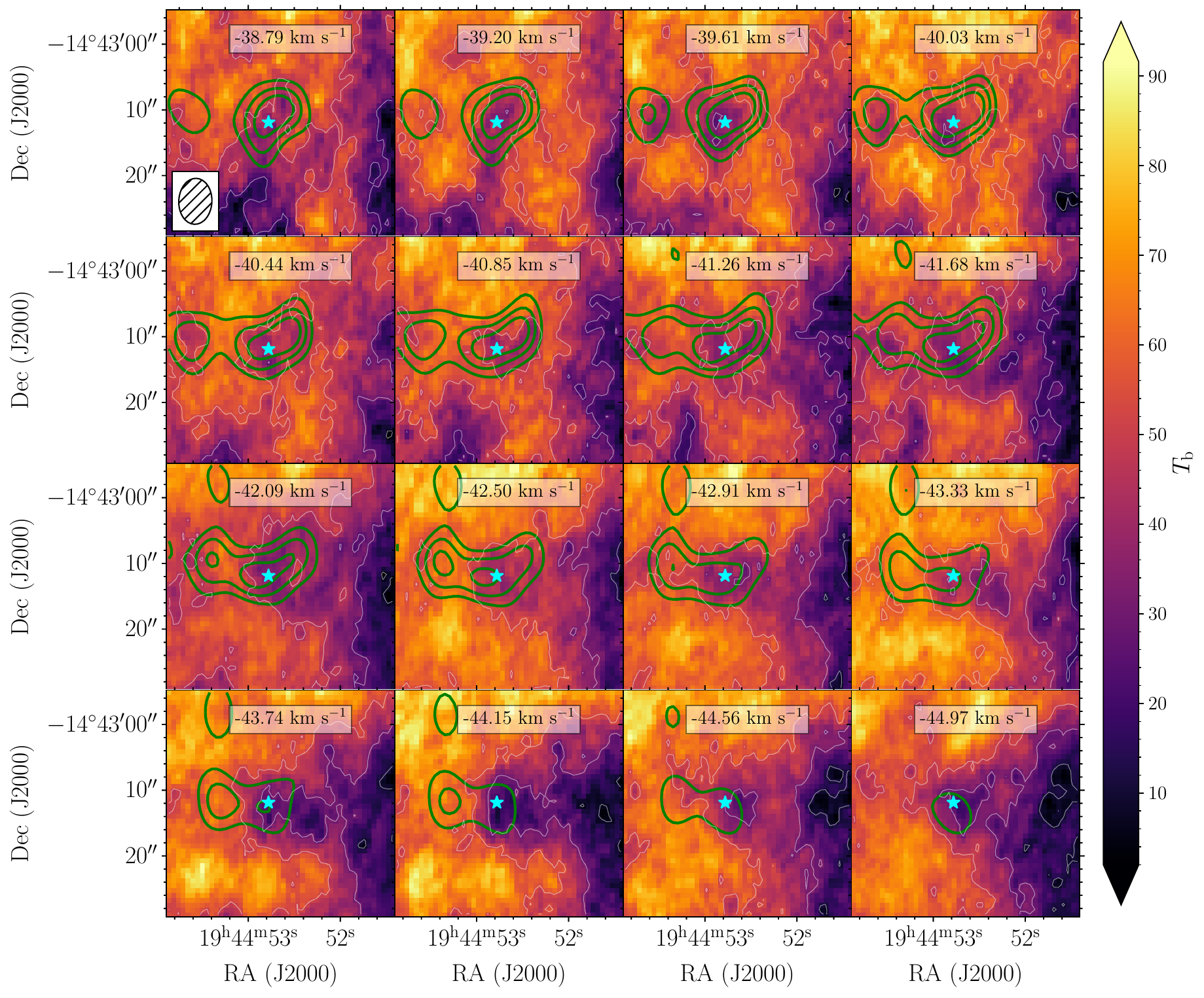}
\caption{\label{fig:HI_CO_chanmaps}Select channel maps of the high-resolution \hi\ emission (background image and white contours) with contours from CO (2$-$1) emission (green) smoothed to the resolution of the \hi\ emission data and the location of the sightline towards NVSS 194452-144311 (cyan star) overlaid. The contours representing \hi\ emission are at levels of 10 K, 30 K, and 50 K, while the CO (2$-$1) emission contours are at levels of 0.25 K, 0.5 K, and 0.75 K.}
\end{figure*}

We expand on the relationship between the atomic and molecular gas phases in this region by showing select channel maps from the high-resolution \hi\ cube in Figure~\ref{fig:HI_CO_chanmaps}, selected to span the velocity range covering the majority of the CO emission. Throughout the entire velocity range, the contours from the CO-bright molecular gas and \hi\ absorption sightline are located within a clear depression of \hi\ emission. 

\begin{figure*}
\includegraphics[width=\textwidth]{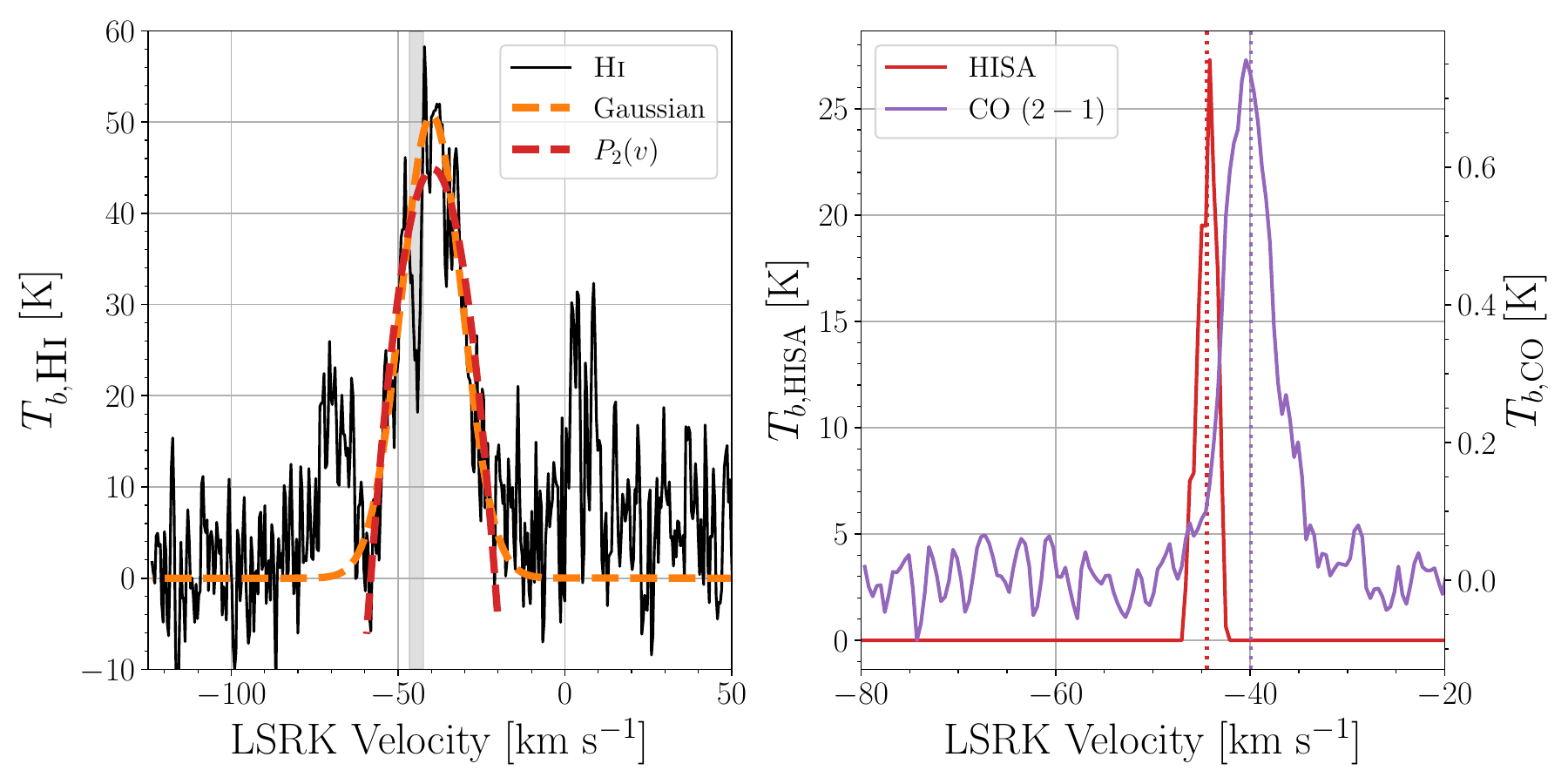}
\caption{\label{fig:HISA_CO} Left: The \hi\ spectrum extracted at $\alpha_{\rm J2000}$=19h44m52.47s,$\delta_{\rm J2000}$=-14$^{\circ}$43m10.90s ($T_{\rm on}$). The narrow dip centered near $-$45 \kms\ results from a potential HISA feature. The orange and red dashed lines respectively represent fits from Gaussian and second-order polynomial functions over the \hi\ emission feature spanning from $-$60 \kms\ to $-$20 \kms\ with velocities from $-$42.5 \kms\ to $-$47 \kms\ excluded, as highlighted by the grey shaded region, to model an off spectrum ($T_{\rm off}$). Right: The HISA spectrum ($T_{\rm off}-T_{\rm on}$), depicted in red (brightness temperature on the left y-axis), compared with a spatially-matched CO ($2-1$) spectrum (brightness temperature shown on the right y-axis). Vertical dashed lines denote the respective intensity-weighted velocity centroids.}
\end{figure*}

\begin{figure}
\includegraphics[width=\columnwidth]{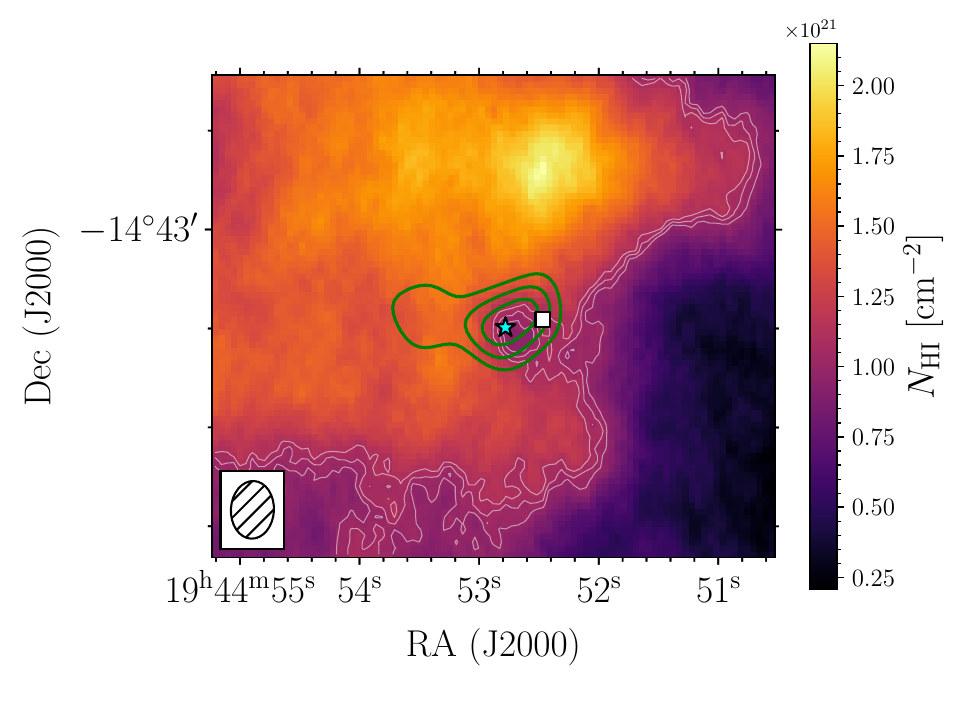}
\caption{\label{fig:HI_H2_colden} A high-resolution \hi\ column density image measured over the $-55$~\kms$<v_{\rm LSRK}<-35$~\kms\ in the optically-thin approximation showing the region around the sightline towards NVSS 194452-144311 (cyan star), with the beam shown in the lower corner. Overlaid in green are contours from the convolved \htwo column density image (see text for details). The faint white contours represent \hi\ column density levels at [1.0, 1.05, and 1.1]$\times10^{21}$ cm$^{-2}$, while the green \htwo column density contours are at levels of [1.0, 2.0, and 3.0]$\times10^{22}$ cm$^{-2}$. The white square marks the location of the spectrum with the HISA feature shown in the left panel of figure~\ref{fig:HISA_CO}.}
\end{figure}


Furthermore, the two highest CO emission contours overlap well with the depression seen in the individual channel maps and are outlined by the lowest \hi\ column density contours. The individual \hi\ spectra within the depression share narrow dips near the systemic velocity of $-44$ \kms for NGC 6822. Given the spatial and spectral association between the \hi\ and CO, we believe this dip results from \hi\ self-absorption (HISA), where foreground CNM attenuates the signal from background WNM. In the left panel of Figure~\ref{fig:HISA_CO}, we show the sightline exhibiting the deepest dip, identified through visual inspection, at $\alpha_{\rm J2000}$=19h44m52.47s,$\delta_{\rm J2000}$=-14$^{\circ}$43m10.90s. This sightline is separated from the one with \hi\ absorption by 4.64$''$. This is almost a full beam width away from the absorption, indicating that the CNM traced by this HISA feature is likely not included in the \hi\ absorption detection. Nevertheless, close spatial and spectral proximity of the HISA and \hi\ absorption suggest they are associated in the overall CNM distribution towards this sightline. The estimated F=0.5 from the Gaussian decomposition analysis (see Section~\ref{fig:gaussian_decomp}) is consistent with the scenerio that at least some WNM is absorbed by foreground CNM in this region. To our knowledge, this is the first instance of HISA observed in an external galaxy outside of the Magellanic Clouds, M31, and M33. 
We follow methods common in analysis of HISA detected in the Milky Way (e.g., \citealt{gibson2005, mcclure-griffiths2006, wang2020}) to estimate the physical properties of the CNM involved in the HISA. In particular, we estimate the optical depth by
\begin{equation}\label{eq:hisa_tau}
    \tau_{\rm HISA}=-ln\left(1-\frac{T_{\rm on}-T_{\rm off}}{T_s-T_c-pT_{\rm off}}\right), 
\end{equation}
where $T_{\rm on}$ is the beam-weighted average \hi\ emission shown in Figure~\ref{fig:HISA_CO}, $T_s$ is the spin temperature of the CNM involved in the HISA, $T_c$ is the background continuum temperature, $p$ is the ratio of the WNM that lies behind and in front of the CNM involved with the HISA, and $T_{\rm off}$ is a model of the \hi\ emission without HISA. We adopt $T_s=27\pm16$ K from the closest CNM component in velocity determined by our radiative transfer decomposition analysis and measure $T_c=78\pm14$ K by taking the mean continuum level over absorption-free velocities in the associated cubelet for the sightline towards NVSS J194452-144311, noting that we apply a gain based on the beam size in the cubelet of 4.85$''\times$3.21$''$. We consider two models for $T_{\rm off}$ by fitting a Gaussian and second-order polynomial ($P_2(v)$) to the broader \hi\ emission feature between $-60$ \kms\ and $-20$ \kms, while excluding the approximate range of the HISA feature extending between $-47$ \kms\ and $-42.5$ \kms. The fraction of background to foreground WNM $p$ is difficult to estimate and generally needs to be assumed based on the properties of the particular sightline. We assume that there is considerable foreground WNM since we are analyzing HISA in an external galaxy and adopt $p=0.1$. The resulting HISA spectrum ($T_{\rm off}-T_{\rm on}$) is shown in the right panel of Figure~\ref{fig:HISA_CO}, using a Gaussian fit to model $T_{\rm off}$. We measure a peak $\tau_{\rm HISA}=0.7\pm0.4$. We estimate the \hi\ column density for the HISA feature using $N_{\hi,\rm HISA}=1.82\times10^{18}T_s\int\tau_{\rm HISA}(v)dv$ as $(6.6\pm4.1)\times10^{19}$ cm$^{-2}$. 
These values are similar to what we found for several CNM structures seen in \hi\ absorption. 
The detection of \hi\ self-absorption shows that cold CNM is even more prevalent in NGC 6822 but that often we are limited by the location of bright background radio continuum sources.

The intensity-weighted velocity centroid of the HISA feature is $-44.4$ \kms, which is offset from the centroid of the spatial resolution-matched CO spectrum shown in the same panel by $-4.5$ \kms. This agrees well with kinematic offsets observed between the CNM in the Milky Way and molecular gas tracers \citep{wang2020, park2022}. Such offsets have been attributed to general relative motions of the atomic and molecular phases set by the local conditions; alternatively, it could be a signature of expansion driven by the strong feedback present in the larger star-forming complex traced by H$\alpha$ emission. 

We check whether this HISA feature results from residual continuum absorption in the high-resolution \hi\ cube by performing an additional image-based continuum subtraction by fitting a 0th order polynomial over the emission-free velocity ranges $-120$ \kms$<v_{\rm LSRK}<-90$ \kms\ and $+30$ \kms$<v_{\rm LSRK}<+50$\kms. We find virtually no changes to estimated properties of this HISA feature. Given the spatial and spectral association with a CO-bright molecular gas complex, we are confident this is genuine HISA signal. 
Furthermore, hints of HISA-like features are seen in individual \hi\ emission spectra that intercept the CO detected by \citet{schruba2017}, though not at a high statistical significance. While we can incorporate the uncertainties for $T_s$ and $T_c$ in our estimate of $\tau_{\rm HISA}$ and $N_{\hi,\rm HISA}$, the variables in equation~\ref{eq:hisa_tau}, such as $p$ and $T_{\rm off}$, must be assumed. But we can quantify the variation in our results by adopting the extreme ends of the allowable values. For example, adopting a second-order polynomial to model $T_{\rm off}$ and setting $p=0.9$ --- i.e., assuming a majority of the observed emission lies behind the absorbing CNM that produces the HISA reduces the estimates for $\tau_{\rm HISA}$ and $N_{\hi,\rm HISA}$ by about a factor of two. Our original Gaussian model for $T_{\rm off}$ and adopted $p=0.1$ results in a peak $\tau_{\rm HISA}$ value that is consistent with the CNM components identified in our radiative transfer decomposition analysis and a $N_{\hi,\rm HISA}$ in agreement with the 3$\sigma$ $N_{\hi}$ threshold of 5.1$\times10^{19}$ cm$^{-2}$ for a 4.5 \kms\ line in the optically thin limit.

\subsection{Molecular fraction and a comparison with theoretical models}

We adopt the CO-to-\htwo conversion factor, $\alpha_{\rm CO}$=85$\pm$25 M$_{\odot}$ pc$^{-2}$ (K \kms)$^{-1}$, for this specific region from \citet{schruba2017} to compare \hi\ and \htwo\ column density distributions over $-55$~\kms$<v_{\rm LSRK}<-35$~\kms\ in Figure~\ref{fig:HI_H2_colden}. We convolve the \htwo\ column density distribution to the beam size of the \hi\ emission (7.0$''\times$5.2$''$) for a more direct comparison. The \htwo\ column density ($N_{\text{\htwo}}$) at the location of the sightline towards NVSS 194452-144311 is $(3.8\pm0.6)\times10^{22}$ cm$^{-2}$. We measure the optical depth-corrected \hi\ column density using the equation (see e.g., \citealt{mcclure-griffiths2023})
\begin{equation}
    N_{\hi,\rm corr}=1.82\times10^{18}\int^{-55~\text{\kms}}_{-35~\text{\kms}}\frac{T_B(v)\tau(v)}{1-e^{-\tau(v)}}dv~ \text{cm}^{-2}
\end{equation}
towards this sightline to be $(2.7\pm0.06)\times10^{21}$ cm$^{-2}$. The $N_{\text{\htwo}}$-to-$N_{\hi,\rm corr}$ ratio is therefore 14.0$\pm$0.2, indicating this sightline is \htwo\-dominated.

\begin{figure}
\includegraphics[width=\columnwidth]{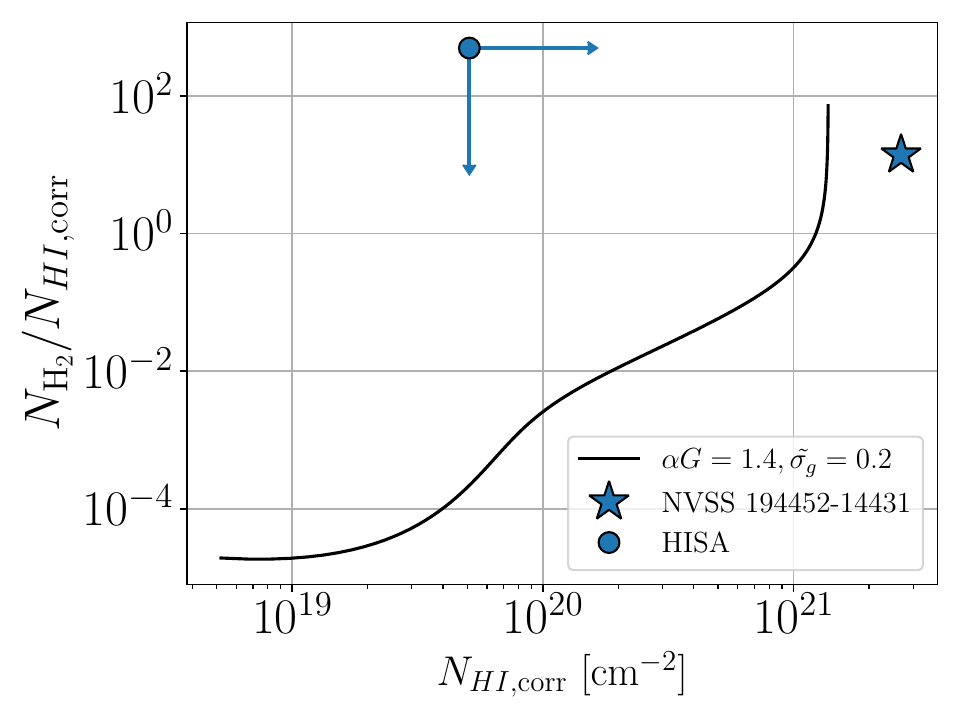}
\caption{\label{fig:H2_to_HI} The observed $N_{\text{\htwo}}$-to-$N_{\hi,\rm corr}$ ratio (blue star) compared with predictions from an analytical model developed in the analytical and numerical works of \citealt{krumholz2008, krumholz2009b, mckee2010, sternberg2014, bialy2016}. Specifically, see the procedure outlined in the introduction of Section 3\citet{bialy2016}. The black curve represent the model prediction with two observationally-constrained input parameters (see text). The error bars for the observed $N_{\text{\htwo}}$-to-$N_{\hi,\rm corr}$ ratio are shown but are too small to see.}  
\end{figure}

We compare our observed $N_{\text{\htwo}}$-to-$N_{\hi,\rm corr}$ ratio with the predictions from analytical models developed in the analytical and numerical works of e.g., \citet{krumholz2008, krumholz2009b, mckee2010, sternberg2014, bialy2016}. 

Two important parameters to consider are the values for $\alpha G$ and $\tilde{\sigma}_g$. The former is physically the ratio of the \hi-dust absorption rate of \htwo\ photodissociation flux to the \htwo\ formation rate. As defined in equation 4 of \citet{bialy2015}, it depends on: the field intensity relative to the mean \citet{draine1978} field, the total hydrogen volume density, the dust-metallicity, and a factor of order unity that depends on the specific grain composition and scattering and absorption properties. The latter parameter is the dust cross-section normalized to 1.9$\times10^{-21}$ cm$^{2}$, which is linearly proportional to the dust-metallicity and the same factor of unity that describes the dust composition. 

We place an observational constraint on $\alpha G$ by measuring the mean UV flux density over the field-of-view of the ALMA CO ($2-1$) observations on an image from the F169M medium-band filter of the Ultraviolet Imaging Telescope (UVIT; \citealt{kumar2012}) aboard the \textit{AstroSat} satellite. We convert this flux density from the calibrated units of \r{A}$^{-1}$ cm$^{-2}$ erg s$^{-1}$ pix$^{-1}$ (where each pixel is 0.42$''\times0.42''$) to 2.5$\times10^{-3}$ MJy/sr. We estimate the mean \citet{draine1978} field to be 1.3$\times$10$^{-8}$ photon s$^{-1}$ cm$^{-2}$ Hz$^{-1}$ sr$^{-1}$, or equivalently 1.6$\times10^{-2}$ MJy/sr using the mean wavelength in the F169M image of 1608 \r{A}. For simplicity, we assume the dust and metal abundance are reduced by the same amount in this region (0.2 Z$_{\odot}$). The steady-state model of \citet{bialy2019} predict a total hydrogen volume density of $\sim$10 cm$^{-3}$ at this metallicity, taking the mean spin temperature of the CNM components as an approximation for the gas kinetic temperature along this line of sight. 

Figure~\ref{fig:H2_to_HI} shows that a model that takes our observationally constrained values for $\alpha G$=1.4 and $\tilde{\sigma_g}=0.2$ into account predicts the \hi\ column density should saturate at a value 60\% lower than our observed ratio. We also place the data point from our nearby HISA detection and note the appropriate lower limit for $N_{\hi,\rm corr}$ and resulting upper limit on the $N_{\text{\htwo}}$-to-$N_{\hi,\rm corr}$ ratio. 

This steady-state analytical model slightly underpredicts the saturation point for the \htwo\-to-\hi\ transition in this sightline. It is difficult to determine which observational constraint and to what magnitude the assumptions in the observational constraints needs to be changed to produce a better match. For example, if $\alpha G$ is ten times higher, \htwo\ is easily photodissociated until about $N_{\hi,\rm corr}\sim10^{22}$ cm$^{-2}$ when the \htwo\-self shielding and dust opacity begins attenuate the photodissociation rate enough for the abundance of \hi\ to saturated. If $\tilde{\sigma}_g$ were smaller, this saturation point will also shift to higher \hi\ column densities due to decreased dust opacity. Similar \htwo\-to-\hi\ ratios need to be measured across a broad parameter space of metallicity and interstellar radiation fields to obtain a better understanding between the observed and predicted saturation of \hi\ column densities that mark the transition to \htwo.


\section{Discussion}\label{sec:discussion}

\subsection{Comparison with Similar \hi\ Absorption Measurements}\label{subsec:comparing_detections}
\begin{figure}
\includegraphics[width=\columnwidth]{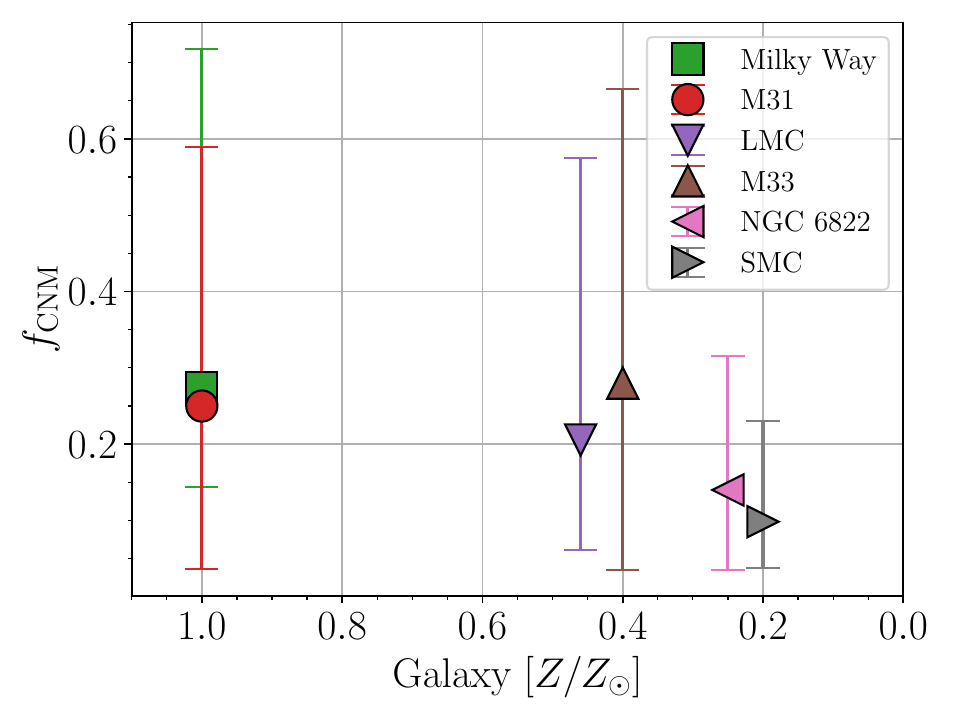}
\caption{\label{fig:f_CNM_vs_metallicity} A scatter plot showing the median $f_{\rm CNM}$ values from individual sightlines through several galaxies as a function of their global $Z/Z_{\odot}$. The error bars represent the 25th and 75th percentiles of the distributions of $f_{\rm CNM}$ for each galaxy. The Milky Way data are taken from the \citet{mcclure-griffiths2023}, with only high-latitude ($|b|>10^\circ$) considered to isolate the local gas to utilize the assumption of solar metallicity. The error bars on this point are smaller than the symbol size. The M31 and M33 points are taken from the integrated quantities summarized in table 3 of \citet{dickey1993}, converting from their assumption that $T_{\rm cold}=60$ K to $T_{\rm cold}=30$ K. Due to sensitivity constraints, the distribution of $f_{\rm CNM}$ values for M31 and M33 are made up of primarily upper limits. The $f_{\rm CNM}$ distribution compiled for the LMC are taken from table 2 of \citet{marx-zimmer2000}, converting $T_{\rm cold}=27$ K to $T_{\rm cold}=30$ K. The SMC values are compiled from \citet{dempsey2022} where we only consider sightlines that have at least one feature above a 5$\sigma$ detection and $<T_{s}>$ values above 100 K. The metallicity values are from the catalogs of \citet{karachentsev2004, karachentsev2013} and measurements from \citet{dominguez-Guzman2022}.}
\end{figure}

We place our measured mass fraction of the CNM for sightlines within NGC 6822 with detections in context with previous absorption measurements in the Milky Way and other nearby galaxies in Figure~\ref{fig:f_CNM_vs_metallicity}. For the sake of comparison and to maintain consistency with values found in the literature, we define $f_{\rm CNM}$ as the fraction of cold gas along the line of sight equal to the ratio of an assumed constant cold gas temperature ($T_{\rm cold}=30$ K) to $\langle T_{\rm s} \rangle$ \citep{dickey1993}, which utilizes the proportionality between temperature and the column density of the CNM and CNM+WNM, respectively. A low-metallicity ISM will not benefit from efficient cooling from fine structure line emission of metals. At the same time, a reduced dust abundance (assuming constant dust-to-metal ratio, which is reasonable for the range of metallicity considered here; \citealt{roman-duval2022}) will decrease the photoelectric heating from PAHs and small grains by FUV \citep{wolfire2003, wolfire2022}. To the first order, the dependence of cooling and heating to metallicity cancels out in the metallicity range explored here ($Z/Z\odot>0.1$; see \citealt{bialy2019,kimjg2023}). The thermal balance between heating and cooling, and hence the CNM fraction, thus depends more subtly on the metallicity via UV radiation transfer and grain charging. Recent numerical simulations of the star-forming ISM including an explicit UV radiation transfer and photochemistry \citep{kimjg2023,kimcg2023} show that the reduced dust attenuation of FUV radiation at low metallicity makes photoelectric heating per stars formed more efficient \citep{kim2024}. For their solar neighborhood models, the resulting $f_{\rm CNM}$ drops from $0.2-0.3$ to $\sim 0.1$ at $Z/Z_\odot=0.3$ and $\sim 0.05$ at $Z/Z_\odot=0.1$ (C.-G. Kim et al. in prep). 
This is indeed the case in the SMC, with a median $f_{\rm CNM}$ around 11\%, while the higher metallicity Milky Way, M31, and M33 possess median $f_{\rm CNM}$ values almost two times as high.  

While our sample of \hi\ absorption detections in NGC 6822 is relatively small compared to the approximately dozen sightlines for M31 and M33, around 200 detections in the SMC, and roughly 1000 sightlines included for the Milky Way, our analysis reveals a qualitative trend. This trend is consistent with theoretical expectations, showing a decrease in $f_{\rm CNM}$ with decreasing metallicity. Our findings demonstrate the type of detailed comparisons we will undertake as we continue to build our sample of \hi\ absorption detections in the LGLBS targets. The LGLBS targets intentionally span a wide range of astrophysical environments (e.g., SFR, metallicity, gas-to-dust ratios) such that we will be able to break degeneracies between trends in CNM properties and local environmental factors that influence its formation, such as strong background UV fields and gas-to-dust content. 


\subsection{Lack of Detections}\label{subsec:lack_of_detections}

Given the similar SFRs and metallicity, it is not surprising that we estimate similar line-of-sight $f_{\rm CNM}$ values (see Section~\ref{subsec:comparing_detections}) between the SMC and the NGC 6822. 
However, our overall detection rate of 11\% (2/18) of \hi\ absorption at the LSRK velocities of NGC 6822 is low in comparison to the SMC where  \citet{dempsey2022} found a detection rate of 38\%. 
As we saw in Section 6.1, there is also some extended cold \hi\ traced by self-absorption that we miss detecting in absorption due to the lack of background radio sources. There are four main reasons for our \hi\ absorption detection rate being lower than that of the SMC.

(1) \textit{A parallel comparison}: The SMC study reported a total detection rate of 38\% across 162 individual sightlines, which were roughly evenly distributed within the 6$\times10^{20}$ cm$^{-2}$ \hi\ column density level. If we consider the same column density threshold to essentially avoid very diffuse areas in the outskirts of NGC 6822, our detection rate increases to 25\% across 8 individual sightlines, becoming much closer to that of the SMC.

(2) \textit{Our continuum observations are slightly less sensitive}: While our median optical depth sensitivity of 0.05 is comparable with \citet{dempsey2022}, the median peak flux density of our background sources is about two times higher at 2.0 mJy/beam as opposed to 0.89 mJy/beam, accounting for the difference in beam sizes. This is expected since we utilize 4 MHz to generate our narrowband continuum image used to build our catalog of background sources, as opposed to the 18.5 MHz bandwidth used to construct the continuum catalog used in \citet{dempsey2022}, meaning we are limited to probing for absorption against stronger sources. 

(3) \textit{NGC 6822 occupies a smaller solid angle}: The solid angle bounded by the 6$\times10^{20}$ cm$^{-2}$ contour in the \hi\ column density image of NGC 6822, generated by integrating over velocities ranging between LSRK velocities of 
$-160$ \kms\ and $-10$ \kms\ to avoid introducing emission from the Milky Way, is just 0.05 deg$^2$. This is two orders of magnitude less than the 21.6 deg$^2$ bounded by the same level in the SMC \hi\ column density image. The proximity of the SMC and a slightly better sensitivity means the GASKAP-\hi\ observations are capable of detecting more suitable background sources over a much larger field-of-view. 

(4) \textit{NGC 6822 has a shorter line-of-sight pathlength}: The 3D structure of NGC 6822 and the SMC is different and this also has implications for the cold HI detection rate. NGC 6822 is more isolated than the SMC, though \citet{deBlok2000} attribute the prominent hole in the \hi\ distribution to the southeast to a potential interaction occurring within the past 100--200 Myr. The SMC is greatly affected by interactions with the LMC and Milky Way, which introduces strong dynamical tidal forces on the gas that will induce more density enhancements through colliding flows to encourage the formation of dense CNM \citep{MAMD2017, tsuge2024}. In addition, the frothy nature of SMC's \hi\ with over 400 expanding shells is conducive to cold cloud formation and displacement throughout the galaxy. 
The SMC is also known to have a large line-of-sight depth, $\sim20$ kpc \citep{kerr1954,johnson1961,hindman1964,murray2024} due to its highly disrupted nature from the large tidal forces applied from the LMC and Milky Way. Therefore, \hi\ absorption in the SMC probes a $\sim$20 kpc total pathlength, providing a significantly higher likelihood of intercepting cold \hi\ for any given sightline. For example, the 60$^\circ$ inclination of NGC 6822 and 0.28 kpc scale height \citep{puche1992, deBlok2000} means we can potentially probe up to a path length of 0.28 kpc/$\sin(i=60^\circ)$=0.32 kpc. Though we note that this scale height is based strictly on dynamical mass arguments as opposed to a more robust direct fit to the \hi\ distribution. Nevertheless, the relatively isolated environment of NGC 6822 compared to the SMC points to this being a reasonable estimate. Assuming that the mean path length between CNM clouds and the filling factor are equal between the SMC and NGC 6822, sightlines in the SMC could intercept up to $\sim$60 times as many individual CNM clouds. 

All of the above reasons suggest that the CNM within NGC 6822 and the SMC have similar local properties, but it is distributed more favourably for detection in the SMC. We will incorporate detection statistics for \hi\ absorption in the other LGLBS galaxies to make a robust quantitative assessment of our detection rate in future work.

\section{Conclusions and Future Work}\label{sec:conculsions}
We presented two detections of \hi\ absorption and \hi\ self-absorption in the low-metallicity, Local Group barred spiral galaxy, NGC 6822 --- the first such detections in a low-metallicity environment outside of the Magellanic Clouds. We summarize our analysis and results and present plans for future work below: 

\begin{itemize}
    \item The Gaussian decomposition and radiative transfer analysis reveals a total of five separate CNM components for the two sightlines. The $T_{s}$ for these components range from $\sim$20 K to 60 K with a mean value of 32$\pm$6 K. We find column densities for the CNM ranging from 1.2$\times$20 cm$^{-2}$ to 8.6$\times$20 cm$^{-2}$; $\Delta v_{\rm FWHM}$ values ranging from 2.8 \kms\ and 4.4 \kms, corresponding to maximum $T_k$ ranging from 180 K to 420 K; and integrated $f_{\rm CNM}$ of 0.16 and 0.37.
    \item We find that our estimated integrated quantities, such as the CNM mass fraction $f_{\rm CNM}$ and the density-weighted mean spin temperature $\langle T_s \rangle$ agree well with \hi\ absorption properties within the SMC, a dwarf galaxy with similarly low metallicity and SFR. We presented a preliminary comparison of the fraction of cold gas along the line of sight$f_{c}$ measured within other nearby galaxies and the Milky Way that demonstrates a slightly decreasing trend between $f_{c}$ with metallicity. A future systematic investigation of $f_{c}$ from other LGLBS galaxies will disentangle the influence from different mechanisms that dictate the thermal phase balance between the WNM and CNM, such as the strength of background UV radiation and dust content.
    \item We stack the \hi\ absorption spectra of 13 non-detections to search for indications of an average low-level signal. We do not see any signatures of the CNM in the stacked profile. We will continue to average down the noise by adding spectra from the other LGLBS dwarf galaxy targets in the future. 
    \item The sightline towards NVSS J194452-144311 intercepts the strongest active star-forming region within NGC 6822, as traced by H$\alpha$ emission. There is good spatial and spectral agreement between the detected CNM, nearby molecular gas traced by CO~(2-1), and the H$\alpha$ emission. Our estimated $N_{\text{\htwo}}$-to-$N_{\hi,\rm corr}$ ratio agrees well with analytical models that accurately predict the saturation of \hi\ at levels$\gtrsim10^{21}$ cm$^{-2}$. This strong link demonstrates that the CNM plays a prominent role in star formation by e.g., shielding star-forming molecular gas from becoming photodissociated.  
    \item We find instances of HISA near the sightline towards NVSS J194452-144311 that is spatially and spectrally correlated with the nearby CO~(2-1) complex. Modeling the \hi\ emission in the absence of HISA by fitting a Gaussian to the broader emission feature, adopting $T_s=27\pm16$K of the nearest CNM component in velocity, and assuming 90\% of the WNM falls in front of the CNM responsible for the HISA,  we estimate $\tau_{\rm HISA}=0.7\pm0.4$, consistent with the optical depth estimates Gaussian decomposition analysis. To our knowledge, this is the first characterization of HISA in an external galaxy beyond the Magellanic Clouds, M33, and M31. While marginal, this HISA signature highlights the power of our combined sensitivity and spectral resolution for LGLBS and provides a unique combination of constraints on the physical properties of the atomic and molecular gas phases of a low-metallicity but actively star-forming ISM outside of Magellanic System. We plan to perform a detailed census of potential HISA in all LGLBS targets in future studies.
\end{itemize}

Our detection rate was 2/18, or 11\%. But this increases to 25\% when focusing on sightlines that fall within the $N_{\hi}$ contour of $6\times10^{20}$ cm$^{-2}$. 
This is slightly lower that the HI absorption detection in the SMC, largely due to the small solid angle (0.05 deg$^{-2}$) of NGC 6822 and the relatively small path length due to the moderate 60$^{\circ}$ inclination. Systematic investigations into the CNM properties of the entire LGLBS sample, which intentionally spans a wide range of astrophysical environments (e.g., star formation, dust content, metallicity) along with \hi\ absorption detected within the Milky Way and Magellanic system from GASKAP-HI, opens up a wide observation parameter space that will enhance our understanding of the local conditions within the atomic ISM that work to set SFR and efficiency. 

\begin{acknowledgments}
We thank the anonymous referee for suggestions that improved the clarity of this paper. This research was supported by the National Science Foundation awards 2030508, 1836650, and 2205630. Support for this research was provided by the University of Wisconsin - Madison
Office of the Vice Chancellor for Research and Graduate Education with funding from the Wisconsin Alumni Research Foundation.
NMP also thanks Dr.~Trey Wenger for useful discussions regarding how to optimally stack non-detection absorption-line spectra.
\end{acknowledgments}

%

\vspace{5mm}
\facilities{The data were imaged using CHTC services: Center for High Throughput Computing. (2006). Center for High Throughput Computing. doi:10.21231/GNT1-HW21}


\software{astropy \citep{2013A&A...558A..33A,2018AJ....156..123A}}



\bibliography{refs}{}
\bibliographystyle{aasjournal}




\end{document}